\documentclass[twocolumn, trackchanges,times,resetfootnote]{aastex701}
\usepackage{xspace}

\newcommand{\eureka}{\texttt{Eureka!}\xspace}
\newcommand{\tiberius}{\texttt{Tiberius}\xspace}
\newcommand{\planet}{HD\,209458\,b\xspace}
\newcommand{\water}{H$_2$O\xspace}
\usepackage[version=4]{mhchem}
\usepackage{bm}
\begin{document}

\title{The asymmetric limbs of \planet observed with JWST~NIRCam~F322W2/F444W}

\author[orcid=0000-0002-9061-780X,sname='Maguire']{Cathal Maguire}
\thanks{These authors contributed equally to this work.}
\affiliation{University of Bristol, HH Wills Physics Laboratory, Tyndall Avenue, Bristol, UK}
\email[show]{cathal.maguire@bristol.ac.uk} 
\author[orcid=0000-0003-0973-8426]{Eva-Maria Ahrer}
\thanks{These authors contributed equally to this work.}
\affiliation{Max Planck Institute for Astronomy (MPIA), K\"onigstuhl 17, 69117 Heidelberg, Germany}
\email[show]{ahrer@mpia.de}  
\author[orcid=0000-0001-9665-5260]{Charlotte Fairman}
\affiliation{University of Bristol, HH Wills Physics Laboratory, Tyndall Avenue, Bristol, UK}
\email{charlotte.fairman@bristol.ac.uk} 
\author[orcid=0000-0003-4328-3867]{Hannah R. Wakeford}
\affiliation{University of Bristol, HH Wills Physics Laboratory, Tyndall Avenue, Bristol, UK}
\email{hannah.wakeford@bristol.ac.uk}
\author[orcid=0000-0002-4997-0847]{Duncan A. Christie}
\affiliation{Max Planck Institute for Astronomy (MPIA), K\"onigstuhl 17, 69117 Heidelberg, Germany}
\affiliation{Department of Physics and Astronomy, Faculty of Environment, Science and Economy, University of Exeter, Exeter EX4 4QL, UK}
\email{christie@mpia.de}
\author[orcid=0000-0002-4207-6615]{James Kirk}
\affiliation{Department of Physics, Imperial College London, Prince Consort Road, SW7 2AZ, UK}
\email{j.kirk22@imperial.ac.uk}
\author[orcid=0000-0002-3052-7116]{Elspeth K.H. Lee}
\affiliation{Center for Space and Habitability, University of Bern, Gesellschaftsstrasse 6, 3012 Bern, Switzerland}
\email{elspeth.lee@unibe.ch}
\author[orcid=0000-0002-9792-3121]{Evert Nasedkin}
\affiliation{School of Physics, Trinity College Dublin, The University of Dublin, Dublin 2, Ireland}
\email{nasedkin@tcd.ie}
\author[orcid=0000-0002-3328-1203]{Michael Radica}
\affiliation{Department of Astronomy \& Astrophysics, University of Chicago, 5640 South Ellis Avenue, Chicago, IL 60637, USA}
\email{radicamc@uchicago.edu}
\begin{abstract}
We present a reanalysis of JWST~NIRCam~F322W2/F444W transit observations of the hot Jupiter \planet to extract morning and evening transmission spectra. To determine whether asymmetries in the transit light curves arise from atmospheric structure, we refine the HD~209458 system parameters through a reanalysis of archival transit and radial velocity data. We then extract limb-resolved spectra using two independent light curve modelling frameworks, \texttt{Harmonica} and \texttt{catwoman}, which yield consistent results. We interpret the transmission spectra using both 1D and 1.5D (i.e., retrievals in which each limb of the planet is modelled independently) atmospheric retrievals. From our 1.5D free-chemistry retrievals, we find a strong preference (${\Delta\ln Z=2.81\pm0.49}$) for a model in which the morning and evening limbs are retrieved independently over one in which they share atmospheric parameters, despite the additional free parameters this requires. This asymmetry is most pronounced in the retrieved cloud properties, consistent with 3D atmospheric circulation models that predict longitudinally varying cloud distributions, and to a lesser extent in the \ce{CH4} and \ce{CO} abundances. In contrast, the \ce{H2O} and \ce{CO2} abundances and the isothermal temperature structure remain largely consistent between limbs. Similarly, our 1.5D equilibrium chemistry retrieval reveals a uniform composition between the limbs, alongside a marginally hotter evening limb relative to the morning limb. While the retrieved temperatures, chemical abundances, metallicities and carbon-to-oxygen ratios ($\rm C/O$) are uniform across both limbs, they exhibit significant differences from our 1D retrieval results, highlighting the importance of multidimensional modelling in mitigating 1D retrieval biases.
\end{abstract}


\section{Introduction} 
%% The "ht!" tells LaTeX to put the figure "here" first, at the "top" next
%% and to override the normal way of calculating a float position.
%% The asterisk after "figure" tells the compiler to span multiple columns
%% if a two column style is selected.
Since the foundational detection of sodium absorption in its atmosphere \citep{charbonneau_detection_2002}, \planet has become one of the most extensively observed exoplanets to date. As the archetypal hot Jupiter, it also often serves as the template to which theoretical models are benchmarked \citep[e.g.,][]{Showman_2009,Rauscher_2013,mayne2014unified}. Owing to its relatively large planet-to-star radius ratio, bright host star ($m_{\rm V}=7.63$),  low density, and large atmospheric scale height, \planet is also among the most amenable exoplanets discovered for characterisation via transmission spectroscopy \citep{Kempton_tsm_2018}.

Multiple transit observations have since studied the atmospheric terminator of \planet in great detail. \cite{Barman_2007_water_stis} first proposed the existence of water vapour in the atmosphere of \planet to explain the excess absorption near $1~\mu$m seen in observations taken with the  Space Telescope Imaging Spectrograph (STIS) aboard the Hubble Space Telescope (HST; \citealt{Knutson_2007_limbdark}). \cite{Beaulieu_2010_water_stis} also detected excess absorption in each of the four photometric channels of the Spitzer InfraRed Array Camera (IRAC), which was again attributed to water vapour. However, a reanalysis of the Spitzer IRAC data presented in \cite{Beaulieu_2010_water_stis}, among others, by \cite{Evans_2015_spitzer} found no evidence for water absorption beyond 3.6$~\mu$m. Furthermore, the extent of this absorption feature, as well as the feature presented in \cite{Barman_2007_water_stis}, was at odds with the amplitude of  the water band resolved by \cite{Deming_2013_wfc3} using HST Wide Field Camera 3 (WFC3). \cite{Deming_2013_wfc3} concluded that a gray opacity source such as a cloud deck was required in order to match the observed water absorption at $1.4~\mu$m, whilst dampening the water feature at $1.15~\mu$m. This conclusion was in agreement with that of \citet{charbonneau_detection_2002}, who found the amplitude of the sodium absorption to be significantly weaker than predicted by clear-atmosphere models, providing the earliest evidence that high-altitude clouds or hazes can mute spectral features in exoplanet atmospheres.

An atmospheric retrieval analysis of the combined HST~STIS, HST~WFC3, and Spitzer~IRAC spectra of \planet also showed evidence of a low altitude gray cloud \citep{Barstow_2017}, with further analysis unveiling a fractional cloud coverage of {$33^{+6}_{-5}$~\%}, with a relatively flat short-wavelength scattering slope indicating the presence of large condensate particles \citep{Barstow_2020}. Additional atmospheric retrieval analysis of the HST~STIS and HST~WFC3 data also showed asymmetric cloud coverage across the planet's terminator, with fractional cloud coverage of {$57^{+7}_{-12}$~\%}, and evidence for a high-altitude cloud deck \citep[$\sim0.01-0.1$~mbar;][]{MacMad_2017,fairman2024importance-0d9}. \citealt{MacMad_2017} also detect nitrogen-bearing species as a source of additional opacity between $1.45\text{--}1.7~\mu\text{m}$.

% Ground-based high-resolution cross-correlation analysis of five transits of \planet taken with the GIANO-B spectrograph detected a myriad of carbon- and nitrogen-bearing species \citep{Giacobbe_2021}, including CO, CH$_4$, C$_2$H$_2$, NH$_3$, and HCN, whilst also confirming the presence of \water. These detections were supported by previous high-resolution secondary eclipse observations of \planet which detected \water, CO, and HCN, on the planet's dayside with CRIRES \citep[CRyogenic high-resolution InfraRed Echelle Spectrograph;][]{Hawker_2018}. However, four transits of \planet observed with the improved CRIRES$^+$ spectrograph only detected \water confidently \citep{Blain_2024}, with non-detections of all other species presented in \cite{Giacobbe_2021}, with the addition of H$_2$S. The non-detection of HCN is attributed to a potential lack of wavelength sensitivity of their CRIRES$^+$ observations ($1.51\text{--}1.78~\mu\text{m}$) to the HCN absorption band centered at {$\sim3.2~\mu$m}, in comparison to the GIANO-B observations ($0.95\text{--}2.45~\mu\text{m}$). Another potential explanation could be the omission of spectral orders which contain significant telluric contamination and/or do not contain significant a cross-correlation signal from the species in question by \cite{Giacobbe_2021}. This has been shown to bias detections by including favourable, spurious cross-correlation signals which might not arise from the existence of absorption lines in the data, but instead from random fluctuations in the cross-correlation function \citep{Cheverall_2023}.

\cite{xue2024jwst-bb8} presented the first transit observations of \planet with the James Webb Space Telescope (JWST), observing one transit each with the NIRCam~F322W2 and F444W filters. They detected both \water and CO$_2$, with evidence of a fractional gray cloud coverage of {$68^{+19}_{-20}~\%$} along the limbs of the planet. Their chemical equilibrium atmospheric retrieval resulted in a 3$\times$ solar metallicity ($[\rm M/H]$) and a sub-solar carbon-to-oxygen ratio (C/O), which they hypothesise is caused by planetesimal enrichment during planetary migration \citep{Oberg_2011}. \cite{Bachmann_2025} confirmed this sub-solar C/O, however found an approximately solar metallicity, by joint-fitting archival HST~WFC3 observations alongside the NIRCam data. This sub-solar C/O ratio could be explained by an incorrect inference of the \water and CO$_2$ abundance from analysing infrared data alone. \cite{Verma_2025} found a solar C/O and $[\rm M/H]$ from equilibrium retrievals of combined optical and infrared observations, including the aforementioned HST~STIS data in their analysis. They show that the inclusion of the optical STIS data alongside WFC3 and NIRCam observations requires comparatively lower abundances of CO$_2$ and H$_2$O, and thus a greater C/O, when compared to analysing the infared data alone. 
% \citep[see also][]{fairman2024importanceopticalwavelengthdata}.
However, with the inclusion of JWST~MIRI/LRS data, \cite{chubb2026magnesiumsilicatecloudsatmosphere} analysed the full $0.6\text{--}12~\mu\text{m}$ transmission spectrum of \planet, and directly detected a magnesium silicate cloud feature. The inclusion of both optical and mid-infrared data allowed for a continuum cloud opacity which truncates the molecular features in the near-infrared, whilst simultaneously aligning their retrieved sub-solar C/O with previous literature values.

The potential presence of inhomogeneous cloud coverage and muted spectral features along the terminator region, combined with the high cadence and sensitivity of JWST~NIRCam observations, motivates a limb asymmetry analysis of \planet. Limb asymmetries describe differences in atmospheric scale height, and thus measured transit depths, between the cooler western (morning) limb and the hotter eastern (evening) limb. This asymmetry is caused by the temperature imbalance in tidally-locked planets between the permanently irradiated dayside and the cooler nightside. This imbalance drives a thermodynamically induced eastward equatorial jet, transporting heat from the dayside to the nightside across the evening limb. Such asymmetries, as well as the eastward equatorial jet, have long been predicted by theoretical General Circulation Models \citep[GCMs;][]{showman2002atmospheric-f14,cooper_showman_2005,parmentier2016transitions-9d2}.  JWST transmission spectroscopy has revealed limb asymmetries in several exoplanet atmospheres to date \citep{Espinoza_2024_wasp39b,Murphy_2024,Murphy_2025,mukherjee2025cloudymorningsclearevenings,fu2025overcastmorningsclearevenings}, spanning equilibrium temperatures of {$\sim770-1540$~K} and exhibiting morning–evening temperature contrasts of {$177-448$~K}. These studies also show a trend of increasing limb temperature contrast with increasing equilibrium temperature. With {$T_{\rm eq}=1450$~K}, \planet lies within this temperature range and squarely within the silicate-dominated cloud regime \citep{Gao_2020}, making it a compelling target for investigating limb asymmetries and their connection to cloud formation across the nightside of hot Jupiters.

This study is structured as follows; in Section~\ref{sec:methods} we outline our methods to reanalyse the JWST observations and revise the system parameters. In Sections~\ref{sec:spec_lc_fitting} \& \ref{sec:atmo_retrievals} we present the light curve fitting using asymmetric transit models followed by the interpretation of the morning and evening spectra using atmospheric retrievals. We then discuss our results in Section~\ref{sec:discuss}. We conclude our study in Section~\ref{sec:conclusions}. Supplementary material is included in the Appendix.

\section{Methods} \label{sec:methods}

\subsection{Analysis of JWST~NIRCam data}
We reanalysed two transits of \planet taken with JWST~NIRCam on November 10 and 16, 2022 (JWST program GTO 1274, PI: J. Lunine). The grism mode time-series mode was used in both cases, with the first transit using the F444W ($3.9\text{--}5.0~\mu\text{m}$) filter and the second using F322W2 ($2.5\text{--}4.0~\mu\text{m}$) filter for the long-wavelength (LW) channel. The short wavelength (SW) channel setup in both cases was  taken with the F212N filter. The analysis and the resulting transmission spectrum was published by \citealt{xue2024jwst-bb8}. For our reanalysis we used two independent reduction pipelines, \eureka and \tiberius.
\begin{figure*}[ht!]
\centering
\includegraphics[width=\linewidth]{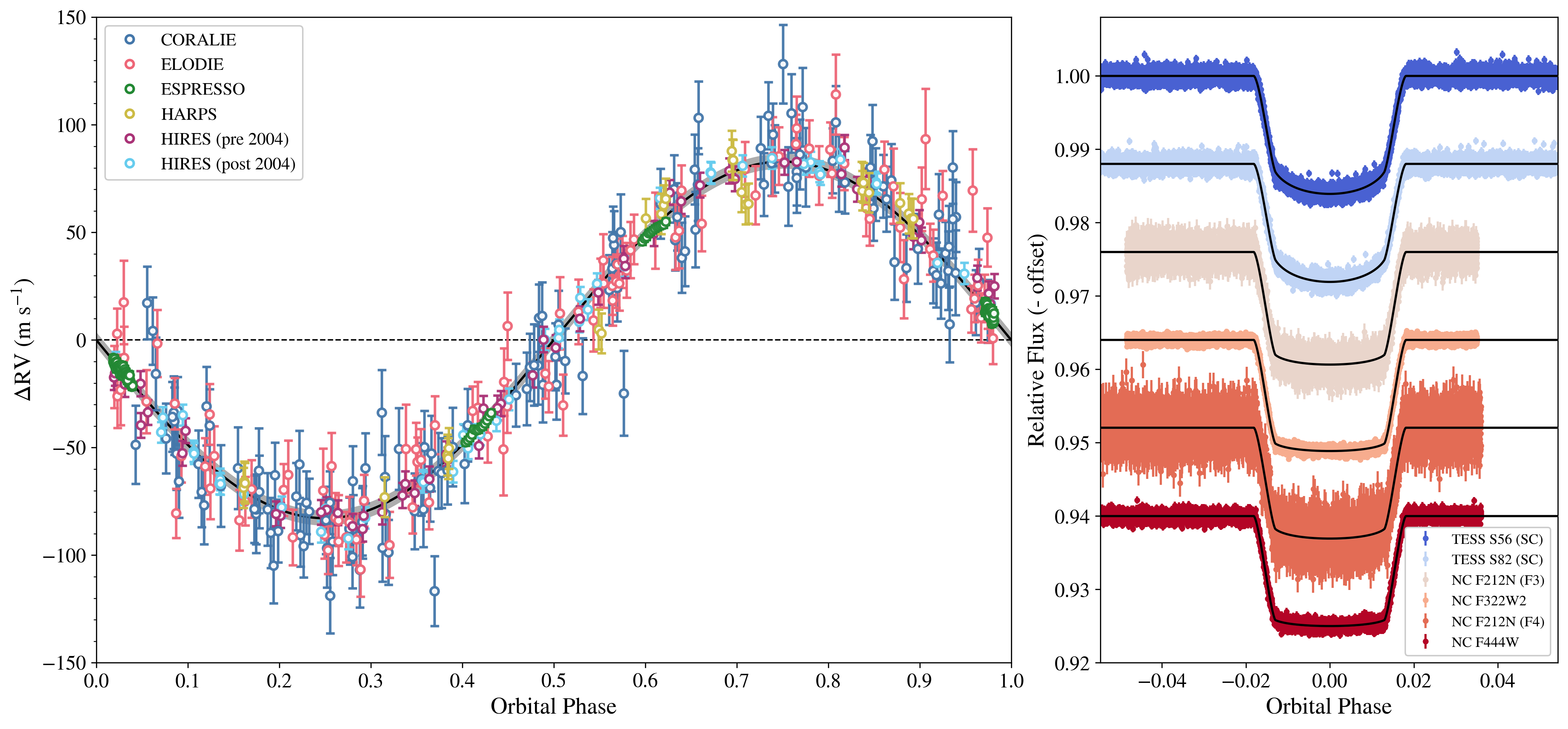}
\caption{(left) The phase-folded radial velocities for the different spectroscopic instruments used in this work. The best-fit Keplerian model for a circular orbit from our \texttt{juliet} fit is shown as a solid black curve, with 1$\sigma$ intervals shaded in grey. The uncertainties shown account for the jitter term, $\sigma_w$. Systemic velocity offsets were applied to bring all RV datasets to a common zero-point, marked by a horizontal dashed line\footnote{The offsets and original datasets are available at \url{https://dace.unige.ch/radialVelocities/?pattern=hd\%20209458}.}. (right) The phase-folded transit light curves for the different instrument used in this work. The best-fit model for a circular orbit from our \texttt{juliet} fit is shown as a solid black curve. SC corresponds to the short-cadence mode of TESS. NC corresponds to JWST~NIRCam. The two short-wavelength NIRCam~F212N light curves are denoted as F3 and F4 to signify their simultaneous acquisition alongside the adjacent long-wavelength F322W2 and F444W observations, respectively}
\label{fig:fig1}
\end{figure*}
\subsubsection{\texttt{Eureka!}}
\label{sec:eureka}
We use \texttt{Eureka!} v1.2.1 \citep{bell2022eureka-6ec} for our first reduction. For both transits in LW channels F322W2 and F444W, we start with running the first two stages of \eureka  which are wrapped around the first two stages of the \texttt{jwst} pipeline \citep{bushouse2024jwst-95d} using the default steps. We only adapt the argument for the rejection threshold in the \texttt{jump} step which we increased to 10.0 and skipped the  \texttt{photom} step. We used the \texttt{jwst\_1364.pmap} CRDS context map for both transit observations. 
We do not include any group-level background subtraction in our fiducial reduction, but we investigate the effect of group-level row-by-row background subtraction (masking the trace) to mitigate potential effects due to 1/f noise. This is particularly challenging for NIRCam data as the readout is in the direction of the trace and thus, 1/f noise manifests in wavelength. We find that including this row-by-row subtraction does not significantly improve the residuals and in fact, minor striping in wavelength remains in the residuals after light curve fitting. We further tested whether including this group-level row-by-row background subtraction affected the final morning/evening transmission spectra, however, we found the results to be consistent (see Figure~\ref{fig:rowbyrow_spectra}). In addition, while noted in another NIRCam data set with similar brightness \citep[HD189733b,][]{fu2024hydrogen-f34}, we do not see the residual 1/f striping change in the amplifier regions, of which NIRCam has four.  

In Stage\,3 of \eureka we extract the time series stellar spectra. We correct for the curvature of the trace (approximately\ 3 pixels over the full wavelength range) by aligning the centre of the trace to the closest integer. We perform weighted average column-by-column background subtraction using the pixels that are $>8$ pixels from the centre of the trace. Afterwards, we use an aperture half width of 3 pixels and optimal spectral extraction \citep{horne1986optimal-8af} to extract the spectrum. 
%For the details on the extraction parameters we refer the reader to the \eureka control files in the Zenodo repository. 

For the two spectroscopic time-series, we use two different binning schemes to be used by the uniform limb fitting versus for the limb asymmetry analysis as the latter requires broader binning to achieve sufficient signal-to-noise. For F322W2 we use 50 and 25 bins equally split across 2.405 --- 4.051 \textmu m for the two cases, respectively. Similarly for F444W we employ 40 and 20 bins in the 3.857 --- 5.056 \textmu m wavelength range for the uniform and limb asymmetric fits, respectively.

We further reduce the two NIRCam SW using \eureka \citep{bell2022eureka-6ec}. Similarly to the LW analysis we use the first two stages of the \texttt{jwst} pipeline \citep{bushouse2024jwst-95d} using the default steps with the exception of the increase of the \texttt{jump} step rejection threshold to 10.0 and skipping the \texttt{photom} step. In Stage\,3 of \eureka we now extract the photometric light curve. We conduct a row-by-row median subtraction. For the  aperture extraction method we use the \texttt{POET} pipeline code originally developed for \textit{Spitzer} photometry \citep{Campo2011WASP-12POET, Stevenson2012POET} implemented within \eureka. In short, it flags bad pixels, calculates the centre from a 2D Gaussian fit and applies aperture photometry. We adopt 35 pixels for the aperture size, and a 20 pixel sky annulus at a distance of 60 pixels from the centre of the trace.

\subsubsection{\texttt{Tiberius}}
\label{sec:tiberius}
For a second reduction, we used \texttt{Tiberius} \citep{kirk2017rayleigh-ae0,kirk2021access-85d} which has been used in multiple JWST studies, including for the reduction of NIRCam data \citep{kirk2024jwstnircam-689}. We began by processing the \texttt{uncal.fits} files through stage 1 of the \texttt{jwst} pipeline (v1.8.2) using the \texttt{jwst\_1464.pmap} CRDS context map. This resulted in \texttt{gainscalestep.fits} files which we then process through \texttt{Tiberius}'s cosmic ray correction algorithm which flags $5\sigma$ outliers within each pixel's time series and replaces these with the median value for that pixel. We then use the \texttt{DQ} mask generated by the \texttt{jwst} pipeline to replace any pixel with a non-zero flag with the median of the nearest, non-flagged, pixels in the cross-dispersion direction. With the cleaned images in hand, we then trace the stellar spectra using Gaussians fitted to every column in the cross-dispersion direction and fit a fourth order polynomial to the means of these Gaussians. We then sum the flux within an 8-pixel-wide aperture after subtracting the background calculated as the median of each column after masking 48 pixels centered on the stellar trace to extract the spectrum.
We bin our spectroscopic light curves following the wavelength grid described in the \texttt{Eureka!} reduction above.

\begin{deluxetable}{lccc}
% \tabletypesize{\normalsize}
\tablewidth{\columnwidth}
\tablecaption{Prior and posterior values for the orbital parameters from the \texttt{juliet} fit of \planet assuming a circular orbit.\label{tab:transit_rv_circ}}
\tablehead{
\colhead{Parameter} & \colhead{Prior} & \colhead{Posterior} & \colhead{Units} 
}
\startdata
\multicolumn{1}{l}{Transit parameters} \\
$P$ & $\mathcal{N}(3.52474859,10^{-6})$ & $3.52474890 \pm 0.00000014$ & days \\
$t_0$ (BJD$_{\rm TDB}-2450000$) & $\mathcal{N}(9893.75125,10^{-4})$ & $9893.751196 \pm 0.000010$ & days \\
$a/R_\star$ & $\mathcal{U}(8.5,9.0)$ & $8.803 \pm 0.011$ & --- \\
$b$ & $\mathcal{U}(0,1)$ & $0.5101 \pm 0.0018$ & --- \\
$i$ & --- (derived) & $86.678 \pm 0.012$ & deg \\
$e$ & 0 (fixed) & --- & --- \\
$\omega$ & 90 (fixed) & --- & deg \\
$R_{\rm{p}}^{\rm TESS~S56~(SC)}$ & $\mathcal{U}(0.1195,0.122)$ & $0.12103 \pm 0.00005$ & $R_\star$ \\
$R_{\rm{p}}^{\rm TESS~S82~(SC)}$ & $\mathcal{U}(0.1195,0.122)$ & $0.12106 \pm 0.00006$ & $R_\star$ \\
$R_{\rm{p}}^{\rm NIRCam~F322W2}$ & $\mathcal{U}(0.1195,0.122)$ & $0.12106 \pm 0.00004$ & $R_\star$ \\
$R_{\rm{p}}^{\rm NIRCam~F322W2~(SW)}$ & $\mathcal{U}(0.1195,0.122)$ & $0.12125 \pm 0.00011$ & $R_\star$ \\
$R_{\rm{p}}^{\rm NIRCam~F444W}$ & $\mathcal{U}(0.1195,0.122)$ & $0.12082 \pm 0.00005$ & $R_\star$ \\
$R_{\rm{p}}^{\rm NIRCam~F444W~(SW)}$ & $\mathcal{U}(0.1195,0.122)$ & $0.12031 \pm 0.00020$ & $R_\star$ \\ \\
\multicolumn{1}{l}{RV parameters} \\
$K_{\rm p}$ & $\mathcal{N}(83.5,5)$ & $82.9 \pm 0.4$ & $\rm m~s^{-1}$ \\
$m$ & $\mathcal{U}(-10^{-2},10^{-2})$ & $-0.0002 \pm 0.0004$ & $\rm m~s^{-1}~day^{-1}$ \\
$c$ & 0 (fixed) & --- & $\rm m~s^{-1}$ \\
\enddata
% \tablenotetext{}{$^*e$ and $\omega$ were fixed to 0 and 90$^\circ$, respectively, for the circular fit. $\mathcal{N}_{[a,b]}(\mu, \sigma)$ denotes a truncated normal distribution with lower and upper bounds $a$ and $b$, and mean and standard deviation $\mu$ and $\sigma$.}
\end{deluxetable}
\subsection{Revision of System Parameters}
Uncertainties in the transit ephemeris, orbital period, or orbital architecture (e.g., inclination, eccentricity) can potentially mimic the effects of limb asymmetries \citep{dobbs-dixon2011impact-fa4,line2016influence-47e,espinoza2024inhomogeneous-db8,Murphy_2024_analytical_asymmetry_time_degen}. Therefore, attributing any asymmetries seen during ingress or egress to variations in the atmosphere relies on accurately known system parameters. We perform a simultaneous fit of archival transit and radial velocity (RV) data of the HD\,209458 system, spanning more than 25 years, using the \texttt{juliet} fitting package \citep{espinoza_2019_juliet}. These include transit observations taken with the Transiting Exoplanet Survey Satellite (TESS) and the JWST~NIRCam~F322W2/F444W white light curve data. We also include the simultaneous NIRCam short wavelength (SW) channel observations taken with the F212N filter. The RV data was taken with the CORALIE, ELODIE, ESPRESSO, HARPS, and HIRES spectrographs \citep{naef_2004_coralie_elodie,wittenmeyer_2005_hires,laughlin_2005_hires,rosenthal_2021,Casasayas_Barris_2021_espresso,barbieri_2023_harps_catalog}, and was acquired using The Data \& Analysis Center for Exoplanets (DACE) database\footnote{\url{https://dace.unige.ch}}.

\indent We analyse a total of 15 individual TESS light curves, comprising 7 transits from Sector 56 and 8 from Sector 82, each observed at the short cadence (SC) of 20~s. Each TESS light curve was trimmed such that they include a baseline of one full transit duration ({$t_{14}=3.072$~hours}) pre- and post-transit (see Figure~\ref{fig:fig1}). The NIRCam data reduction is outlined in Section~\ref{sec:eureka}. For each instrument, we calculated and held fixed the four parameter non-linear limb-darkening coefficients \citep{claret_2000_4param_LDC} with \texttt{ExoTiC-LD} \citep{Grant2024_exoticLD} using the three-dimensional Stagger stellar model grid \citep{magic2015stagger}. \\
\indent To avoid any potential contamination by the Rossiter-McLaughlin effect, in-transit radial velocity measurements were flagged and removed from our fitting routine. In August 2004, the HIRES spectrograph underwent a significant upgrade \citep{butler_2017_hires}. Consequently, radial velocity measurements obtained prior to August 2004 (pre~2004) and after August 2004 (post~2004) were treated as independent datasets, each with its own RV offset ($\mu$) and jitter term ($\sigma_w$). We fix the y-intercept of the long-term linear trend in the radial velocities, $c$, to 0, to prevent degeneracy with the instrument-specific RV offsets. An overview of all fixed and fitted parameters used in the joint transit and RV fit, as well as their prior and posterior values, is provided in Tables~\ref{tab:transit_rv_circ} and \ref{tab:transit_rv_inst}. We typically adopt uniform priors for each free parameters, however for the orbital period, $P$, and transit ephemeris, $t_0$, we use Gaussian priors informed by the posterior distributions reported in \cite{xue2024jwst-bb8}. 

The best-fit models for the light curves and phase-folded RVs are shown in Figure~\ref{fig:fig1}, respectively. Although we initially modelled a circular orbit, we specifically investigated the impact of eccentricity given its potential to mimic limb asymmetries \citep{Espinoza_2024_wasp39b}. Our eccentric fit found ${e=0.0112\pm0.0028}$ and ${\omega=62.1^{+10.7}_{-7.5}~^\circ}$, where $e$ is the orbital eccentricity and $\omega$ is the argument of periastron. To test the robustness of our results, we repeated the spectroscopic light curve fits (Section~\ref{sect:harmonica_fitting}) using these eccentric parameters (see Table~\ref{tab:transit_rv_ecc}), however increasing the eccentricity by 3$\sigma$, where $\sigma$ is the standard deviation of $0.0028$. We found the recovered morning/evening spectra to be consistent with those obtained using the system parameters from our circular fit. Furthermore, we verified that our orbital solutions, particularly the circular case, are consistent with the Spitzer secondary eclipse times from \cite{evans2015uniform-8de}. Therefore, we opt to keep our system parameters fixed to the circular fit values shown in Table~\ref{tab:transit_rv_circ} for the remainder of the study.
\section{Spectroscopic light curve fitting} \label{sec:spec_lc_fitting}
To test the robustness of our extracted limb spectra, as well as compare two commonly used asymmetric light curve fitting algorithms in the literature, we opt to fit our spectroscopic light curves with both the \texttt{Harmonica} \citep{grant2023harmonica} and \texttt{catwoman} \citep{Jones2021Catwoman:Curves,espinoza2024inhomogeneous-db8} packages. \texttt{Harmonica} models the asymmetric terminator as a ``transmission string'', whose effective radius, $R_{\rm p}$, varies as a function of limb angle, and is defined by a Fourier series of harmonic order $N_c$. In contrast, \texttt{catwoman} parameterises the asymmetric terminator as two back-to-back semicircles, each defined by a distinct radius ($R_{\rm p}^{\rm Eve}$ and $R_{\rm p}^{\rm Morn}$).
\subsection{Spectroscopic light curve fitting with \texttt{Harmonica}}
\label{sect:harmonica_fitting}
The NIRCam~F322W2/F444W spectroscopic light curves were first modelled using \texttt{Harmonica}. This allows the planet's effective radius to vary as a function of limb angle, $\theta$, relative to the morning equator. In order to directly compare \texttt{Harmonica} to the back-to-back semicircle approach of \texttt{catwoman} (Section~\ref{sec:catwoman_fitting}), we restrict the Fourier series to the first harmonic order ({$N_c=1$}). Furthermore, we set the amplitude of the sine term ($b_1$) to $0$, such that
\begin{equation}
\label{eq:harmonica}
    R_{\rm p}(\theta)~=~a_0 + a_1\cos(\theta)
\end{equation}
where $a_0$ is the base planetary radius and $a_1$ is the amplitude of the cosine term of the first harmonic order. This reparameterisation ensures the transmission string is symmetric about the equator, and any variation in the effective radius arises solely from differences between the morning and evening limbs (see Figure 2 of \citealt{grant2023harmonica}). \\ \indent We model the observed light curve as the product of a transit signal, $T(t,\Theta)$, computed using \texttt{Harmonica}, and a linear systematics component, $S(t_s)$, such that
\begin{equation}
\label{eq:harmonica_transit_model}
M(t)~=~T(t,\Theta)\times S(t_s)
\end{equation}
where $\Theta$ denotes the parameter vector describing the transit model, and $t_s$ represents the time relative to the observation start time. Due to a degeneracy between the central transit time, $t_0$, and $a_1$, we choose to marginalise over both parameters in our light curve fitting. We also fit for $a_0$, as well as a linear systematics model given by
\begin{equation}
\label{eq:sys_model}
S(t_s)~=~c_0 + c_1 t_s
\end{equation}
where $c_0$ and $c_1$ are constants to be fit. We also allow our flux uncertainties to be inflated by a factor $\beta$. As before, for each wavelength bin we calculated and held fixed the four parameter non-linear limb-darkening coefficients \citep{claret_2000_4param_LDC} with \texttt{ExoTiC-LD} \citep{Grant2024_exoticLD} using the three-dimensional Stagger stellar model grid \citep{magic2015stagger}.
% When fitting for limb asymmetries, it is essential that we account for delays in the readout time across the detector for each of our wavelength bins \citep{alderson_wasp17b}. As the NIRCam's grism mode disperse light horizontally across our detector, the timestamp of each exposure varies slightly with position, and thus wavelength. However, the readout pattern occurs vertically along the detector, thus any variations in readout time along the trace position, and thus wavelength, are negligible.
\subsection{Spectroscopic light curve fitting with \texttt{catwoman}}
\label{sec:catwoman_fitting}
To extract the morning and evening spectra of \planet from the spectroscopic light curves we also employ the \texttt{catwoman} package. \texttt{catwoman} assumes that the exoplanet is split into two semi-circles, with the angle $\phi$ describing the angle of rotation of the top semi-circle
with regard to the orbit direction, i.e.,  we set {$\phi=90^\circ$} such that the top semi-circle is equal to the evening terminator side.
\begin{figure*}[ht!]
\includegraphics[width=\linewidth]{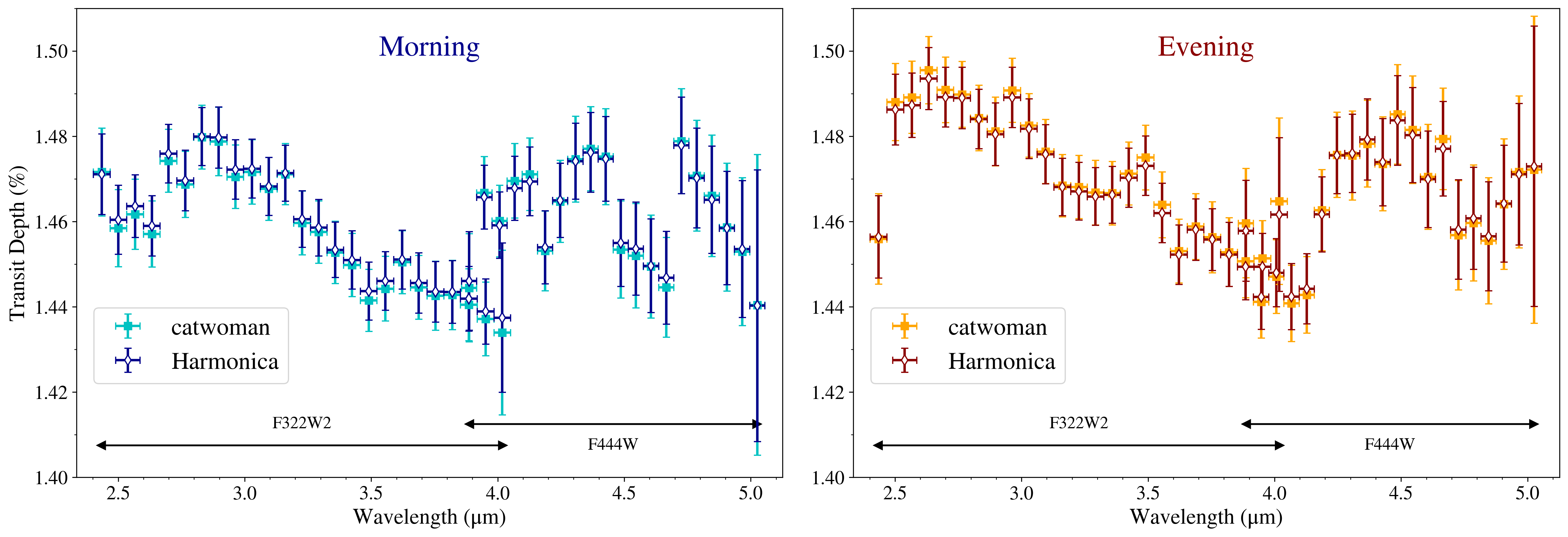}
\caption{The morning (left) and evening (right) transmission spectra obtained for each of our light curve fitting routines (detailed in Sections~\ref{sect:harmonica_fitting}~\&~\ref{sec:catwoman_fitting}). The light curves were reduced with \texttt{Eureka!} (see Section~\ref{sec:eureka}). The transit depths obtained from the \texttt{Harmonica} fit are shown as unfilled diamonds, whereas the transit depths obtained from the \texttt{catwoman} fit are shown as filled squares. Horizontal black arrows highlight the wavelength coverage of the NIRCam~F322W2 and F444W filters. The best-fit offset value from our 1.5D fiducial free-chemistry atmospheric retrieval has been applied to the F444W values for clarity.}
\label{fig:catharm_eureka}
\end{figure*}
Similar to \texttt{Harmonica}, we model the observed light curve as the product of a transit signal, $T(t,\Theta)$, and the linear systematics component, $S(t_s)$, see Equation\,(\ref{eq:harmonica_transit_model}). This time, $T(t,\Theta)$ is computed using \texttt{catwoman}, which consists of the two semi-circles described by $R_{\rm p}^{\rm Eve}/R_\star$ and $R_{\rm p}^{\rm Morn}/R_\star$, the evening and morning planet to stellar radius ratio, respectively. We also allow our flux uncertainties to be inflated by a factor $\beta$. The limb-darkening is described by the four parameter non-linear law, using the same generated coefficients as \texttt{Harmonica}. Thus, for every spectroscopic light curve we fit six free parameters ($t_0$, $R_{\rm p}^{\rm Eve}/R_\star$, $R_{\rm p}^{\rm Morn}/R_\star$, $c_0$, $c_1$, and $\beta$).

\section{Atmospheric retrievals}
We perform two independent atmospheric retrieval analysis to determine which atmospheric parameters (composition and/or clouds) may explain the differences in the morning and evening spectra. We describe the setups for both below.

\label{sec:atmo_retrievals}
\subsection{Atmospheric Retrievals with \texttt{Exo Skryer}}
\label{sec:exo_skryer_retrievals}
We used the atmospheric modelling and retrieval code \texttt{Exo Skryer} \citep{Lee_2026_exoskryer} to infer the atmospheric properties of both the morning and evening limbs of \planet. We built a 1.5D atmospheric model in which the transmission spectra of both limbs were modelled and retrieved simultaneously. This 1.5D modelling approach allows the limbs to share certain parameters whilst allowing others to remain independent, as outlined below. We compared each of our models to a 1D atmospheric model, whose parameters were inferred from transit depths obtained from a uniform (i.e., $N_c=0$) light curve fit. We used the nested sampling algorithm \texttt{dynesty} \citep{speagle2020dynesty-00a} to sample the parameter space, adopting 1500 live points. A standard Gaussian log-likelihood was used to compare each forward model to its corresponding observed transmission spectrum. The atmospheric retrievals were performed on both the \texttt{Harmonica} and \texttt{catwoman} spectra.

The forward model consisted of 99 pressure layers equally distributed in log space from 10$^{-8}$ to $10^2$\,bar, with an isothermal vertical temperature-pressure ($T$--$P$) profile. We employed both a free-chemistry and equilibrium chemistry scheme, outlined below. A gray (i.e., wavelength-independent) cloud profile was used, whose opacity $\kappa_{\rm cld}$ decays exponentially with decreasing pressure above the cloud base. The vertical cloud structure is parameterised by a base cloud pressure, $P_{\rm base}$, and a decay parameter, $\alpha_{\rm cld}$. We also included Rayleigh scattering from H$_2$ \citep{Bell_1980_h2_opacity} and He \citep{Bell_1982_he_opacity} as well as collision-induced absorption (CIA) from both {H$_2$-H$_2$} and {H$_2$-He} \citep{richard2012new-ff0}. For our 1.5D retrievals, we treated the parameters governing the global geometry and deep atmospheric structure as shared between both limbs---specifically the planetary radius ($R_{\rm p}$), surface gravity ($g$) and reference pressure ($P_{\rm ref}$), similarly to the 1.5D modelling approach outlined in \cite{mukherjee2025cloudymorningsclearevenings}. See Tables~\ref{tab:free_chem} \& \ref{tab:chem_eq} for an overview of the atmospheric parameters and their associated priors used in each of our 1.5D \texttt{Exo Skryer} atmospheric retrievals.
\subsubsection{Free-chemistry retrievals}
\label{sec:freechemretrieval}
To infer the presence of any molecular species in our morning/evening spectra, we initially allowed a constant molecular volume mixing ratio (VMR) with altitude. We included molecular opacities from H$_2$O \citep{polyansky2018exomol-f3b}, CO$_2$ \citep{yurchenko2020exomol-ce6}, CH$_4$ \citep{yurchenko2024exomol-63d}, and CO \citep{Li2015_co_crosssections}, utilising correlated-$k$ tables ($R=250$) for the cross-sections. For our free-chemistry retrievals, we compared two distinct model setups. In the first run, the temperatures were fit completely independently between the morning and evening limbs. In the second run, the limb temperatures were linked via:
\begin{equation}
\label{eqn:linked_temps}
  T_{\rm iso}^{\rm Morn}~=~T_{\rm iso}^{\rm Eve}~-~\Delta T_{\rm iso}^{\rm Morn}
\end{equation}
 where $\Delta T_{\rm iso}^{\rm Morn}$ is a temperature offset parameter whose prior distribution was a uniform distribution bounded between 0~K and 500~K. This ensured the morning limb temperature did not exceed the evening limb temperature, akin to the modelling approach presented in \cite{mukherjee2025cloudymorningsclearevenings}.
 
\subsubsection{Equilibrium chemistry retrievals}
\label{sec:equilchemretrieval}
We also performed chemical equilibrium retrievals on the morning and evening transmission spectra to infer the atmospheric carbon-to-oxygen ratio ($\rm C/O$) and metallicity ($[\rm M/H]$) of each limb. We use the same set of gaseous species as in our free-chemistry retrievals. For these retrievals, the chemical equilibrium abundances were calculated using \texttt{FastChem} \citep{Stock_2018}, adopting solar elemental abundances from \citet{asplund_2021}. The metallicity is parameterised as a uniform scaling of all elements heavier than helium relative to the adopted solar composition, with $[\mathrm{M/H}]$ representing the logarithmic ratio of the atmospheric metallicity to the solar value. The C/O ratio is then adjusted by modifying the carbon abundance while keeping oxygen fixed at its metallicity-scaled value. Similarly to the free-chemistry retrievals outlined above, we performed two runs, one with a shared C/O and [M/H] across the limbs, and one where C/O and [M/H] were allowed to vary independently between the limbs. 

\subsection{Atmospheric Retrievals with \texttt{petitRADTRANS} using equilibrium chemistry}
\label{sec:prt_retrievals}
We ran additional 1.5D chemical equilibrium retrievals using the open-source radiative transfer code \texttt{petitRADTRANS} v3.1 \citep{mollire2019petitradtrans-7c8,blain2024spectralmodel-d73, nasedkin2024atmospheric-ed7} to validate the \texttt{Exo Skryer} results. We also ran a 1D retrieval with our \eureka data combined with the available HST data sets. The gaps in time-series HST observations do not allow the cadence and continuous phase coverage required for limb-asymmetry analyses, and, by extension, 1.5D retrievals. We describe the \texttt{petitRADTRANS} setup in the following paragraphs, while the individual 1.5D and 1D retrieval details are described in the subsections.

In both cases, we used the \texttt{MultiNest} implementation \citep{feroz2009multinest-e8d, buchner2014x-ray-2b1}, using 1000 live points for the JWST-only 1.5D retrievals, and 1300 live points for the JWST+HST 1D retrieval. We assumed equilibrium chemistry using the implemented pre-calculated chemistry table from \texttt{easyCHEM} \citep{lei2025easychem-d47}, first described by \citet{mollire2017observing-793}. In contrast to \texttt{FastChem} (Section~\ref{sec:equilchemretrieval}), \texttt{easyCHEM} adjusts the C/O ratio by modifying the oxygen abundance after scaling all elements heavier than hydrogen and helium according to the metallicity.

We set our reference pressure to 0.1\,bar and used 100 atmospheric layers equally distributed in log space from 10$^{-8}$ to $10^2$\,bar. We assumed a separate isothermal $T$--$P$ profile for each limb. 
For the opacities, we use \texttt{petitRADTRANS}'s correlated-$k$ opacity tables at $R=1,000$. We included \ce{H2O} \citep{polyansky2018exomol-f3b}, \ce{CO2} \citep{yurchenko2020exomol-ce6}, \ce{CO} \citep{rothman2010hitemp-614}, \ce{CH4} \citep{hargreaves2020accurate-7e7},  \ce{HCN} \citep{barber2014exomol-f1e}, \ce{H2S} \citep{azzam2016exomol-aa6}, and \ce{C2H2} \citep{chubb2020exomol-d41}. We further considered \ce{H2-H2} and \ce{H2-He} collision-induced absorption \citep{richard2012new-ff0}.
We fitted for an opaque layer at a cloud deck pressure $\log P_\mathrm{cloud}$ and fractional cloud factor $\phi$. Additional scattering parameters and opacities are added for the 1D retrieval and described in Section\,\ref{sec:jwst_hst_rets}.
We assumed a stellar radius of 1.2 R$_\odot$.
All prior distributions together with the posterior median values and their respective uncertainties are shown in Table\,\ref{tab:prt_priors}.

\subsubsection{1.5D equilibrium chemistry retrieval}
We built an atmospheric model similar to \texttt{Exo Skryer} to investigate potential differences and/or biases between the atmospheric retrieval codes.  We shared the planet parameters (planetary mass and radius) between morning and evening, as well as the chemistry parameterised by the C/O ratio and [M/H] of the atmosphere. We also fitted for an offset between F322W2 and F444W. The differences in the limb spectra are then based on the temperature ($T_\mathrm{iso}^{\rm Morn}$, $T_\mathrm{iso}^{\rm Eve}$) and cloud-top pressure ($\log P_\mathrm{cloud}^{\rm Morn}$, $\log P_\mathrm{cloud}^{\rm Eve}$).

\subsubsection{Uniform 1D retrieval JWST + HST}
\label{sec:jwst_hst_rets}
We performed a retrieval analysis using the uniform 1D spectrum of our reanalysed \textit{JWST} spectrum combined with the \textit{HST} spectra from \cite{sing2016continuum-9d7}. This provided a comparison to the retrieval results from \citet{Bachmann_2025, Verma_2025, chubb2026magnesiumsilicatecloudsatmosphere} which used the \textit{JWST} spectrum from \citet{xue2024jwst-bb8} and the \textit{HST} data. 

As the wavelength range extends into the optical, we included the absorption features of broadened Na \citep{allard2019new-b2f} and K \citep{allard2016kh2-1c6}, as well as scattering opacity in our model. We parameterised the latter in the form of a power-law such that the following is added to the scattering cross-section:
\begin{equation}
    \kappa = \kappa_0 \left( \frac{\lambda}{\lambda_0}\right) ^\gamma
\end{equation}
where $\kappa_0$ is the opacity at wavelength $0.35$~\textmu m in units of cm$^2$/g, and $\gamma$ describes the wavelength dependence (with a Rayleigh-like scattering at $\gamma=-4$).

The best-fit model together with the opacity contributions and the JWST + HST data sets are shown in Figure \ref{fig:prt-1d-contributions}.

\begin{figure*}[ht!]
\includegraphics[width=\linewidth]{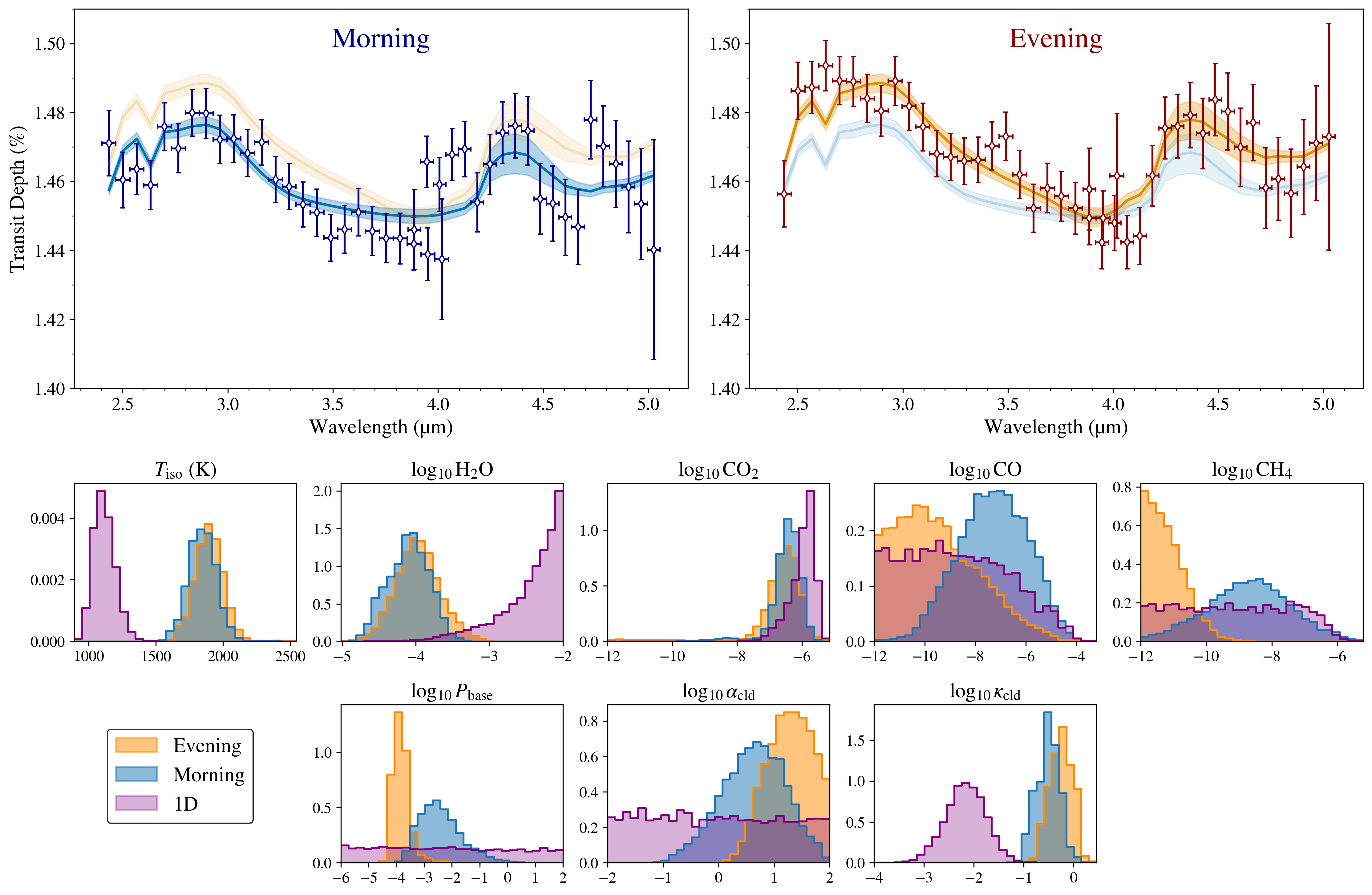}
\caption{Results of the fiducial 1.5D \texttt{Exo Skryer} free-chemistry retrieval (see Section~\ref{sec:freechemretrieval}) applied to the \texttt{Harmonica} transmission spectra. The top panels display the median model transmission spectra and 1$\sigma$ confidence intervals derived from 10,000 posterior samples, with the morning and evening spectra shown on the left and right, respectively. For reference, the evening and morning median models are also shown with a lower opacity on the left and right, respectively. The bottom panels present the marginal posterior distributions for the corresponding limb parameters, with the morning and evening posteriors highlighted in blue and orange, respectively. For comparison, the marginal posteriors from our 1D NIRCam-only atmospheric retrieval are plotted in purple.}
\label{fig:freechem_retrieval}
\end{figure*}
\begin{figure*}[ht!]
\includegraphics[width=\linewidth]{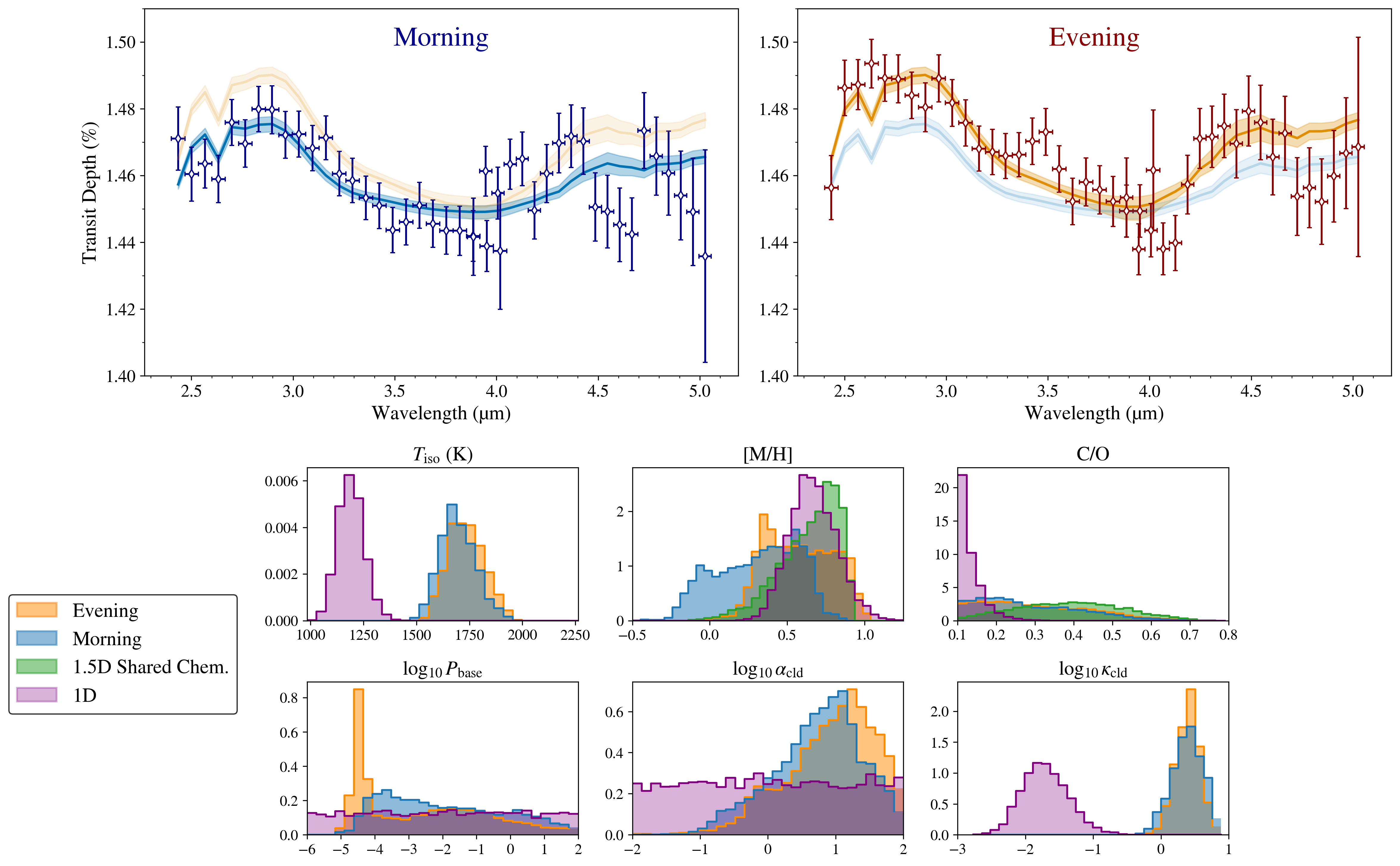}
\caption{Results of the fiducial 1.5D \texttt{Exo Skryer} equilibrium chemistry retrieval (see Section~\ref{sec:equilchemretrieval}) applied to the \texttt{Harmonica} transmission spectra. The top panels display the median model transmission spectra and 1$\sigma$ confidence intervals derived from 10,000 posterior samples, with the morning and evening spectra shown on the left and right, respectively. For reference, the evening and morning median models are also shown with a lower opacity on the left and right, respectively. The bottom panels present the marginal posterior distributions for the corresponding limb parameters, with the morning and evening posteriors highlighted in blue and orange, respectively. For comparison, the marginal posteriors from our 1D NIRCam-only atmospheric retrieval are plotted in purple. Similarly for the [M/H] and C/O posteriors, the marginal posteriors from a 1.5D atmospheric retrieval assuming shared chemistry between the limbs are plotted in green.}
\label{fig:equilchem_retrieval}
\end{figure*}

\section{Discussion} \label{sec:discuss}
\subsection{Spectroscopic light curve fitting}
The NIRCam~F322W2/F444W spectroscopic light curves were modelled using two separate light curve fitting routines, \texttt{Harmonica} and \texttt{catwoman}. The details of each fitting routine are outlined in Sections~\ref{sect:harmonica_fitting} \& \ref{sec:catwoman_fitting}, respectively.

For the \texttt{catwoman} fits, the morning and evening transit depths, and their uncertainties, were computed directly from these morning and evening radii posterior samples. To calculate the morning and evening transit depths from the \texttt{Harmonica} fits, we generated a transmission string for each posterior sample and computed the median radius over the angular ranges corresponding to the morning and evening limbs (i.e., {$0–90^\circ$} and {$270–360^\circ$} for the morning limb, and {$90–270^\circ$} for the evening limb). The associated uncertainties were calculated from the standard deviation of these radii across the posterior samples. We find that this approach yields transit depth uncertainties consistent with those recovered from the \texttt{catwoman} fit, in contrast to computing the radii/uncertainties solely at 0$^\circ$ and 180$^\circ$ for the morning and evening limbs, respectively. The resulting morning and evening transmission spectra for both the \texttt{Harmonica} and \texttt{catwoman} fitting routines, for the \texttt{Eureka!} reduced light curves, are shown in Figure~\ref{fig:catharm_eureka}. Despite the differences in parameterisation and transit depth calculation, we see excellent agreement between both light curve fitting routines. These spectra are also robust to reduction choice (see Figure~\ref{fig:catharm_tiberius}). The individual morning and evening spectra for each fitting routine are also plotted in Figure~\ref{fig:morneve_comp}.

By restricting \texttt{Harmonica} to the first harmonic order and setting the amplitude of the sine term to zero, any east--west asymmetry is modelled as a global shift along the equator. This inevitably links the morning and evening radii, meaning any decrease in the morning radius must result in an increase in the evening radius, and vice versa. This can be seen in the slight anti-correlation between the \texttt{Harmonica} morning/evening transit depths in Figure~\ref{fig:morneve_comp}. A similar anti-correlation can also be seen in the \texttt{catwoman} morning/evening spectra.

To investigate the impact of this anti-correlation on our 1.5D retrievals, we incorporated the covariance between the morning and evening transit depths into our likelihood, following the approach adopted in previous limb-asymmetry retrieval analyses \citep{espinoza2024inhomogeneous-db8}. However, we found that the correlated likelihood did not produce stable, physically meaningful retrievals. We therefore adopt independent likelihoods for our 1.5D retrievals, neglecting the covariance between the morning and evening transit depths, in order to recover meaningful variations in the retrieved parameters.

% However, as the covariance term weights the model on the morning--evening difference at each wavelength, the retrieval became more sensitive to reproducing this difference rather than fitting either spectrum individually. Thus, the sampler drove the two limbs towards opposing extremes, flattening the morning spectrum via a high-altitude cloud deck and a low temperature while simultaneously inflating the evening spectrum, in order to reproduce the observed morning--evening residuals (see Figure~\ref{fig:pRT_cov}). 
\subsection{Excess absorption on the evening limb at 3.5 $\mu$m}
We report an excess absorption feature between \mbox{$3.4-3.6$~$\mu$m} on the evening limb spectrum of \planet (see Figures~\ref{fig:catharm_eureka} \& \ref{fig:morneve_comp}), which may indicate the presence of an unmodelled atmospheric species. However, we note a slight anti-correlation with this potential feature on the morning limb across this wavelength range, and thus do not rule out potential systematics as the cause of the increase in apparent radius. As shown in Figures~\ref{fig:freechem_retrieval} \& \ref{fig:equilchem_retrieval}, this potential feature is not fit by any gaseous species in our atmospheric models. This spectral window contains many prominent absorption features, most notably C$-$H vibrational stretching modes characteristic of complex hydrocarbons \citep{Niraula_2025}. A preferential excess of complex hydrocarbons on the evening limb is intriguing, as we would expect photochemical production of hydrocarbon hazes due to intense stellar irradiation on the dayside, followed by day--night advection. Simulations of photochemical haze transport in the atmosphere of \planet have shown that these haze particles preferentially accumulate instead over the morning limb \citep{mak_2025_photohazes}. \citet{Richardson_2007} reported a tentative feature at $7.78$~$\mu$m in the Spitzer InfraRed Spectrograph (IRS) emission spectrum of \planet, which they noted could be consistent with the C$-$C stretching resonance of polycyclic aromatic hydrocarbons (PAHs). PAHs also exhibit a prominent absorption feature near $3.3$~$\mu$m \citep{Dubey_2023}, which also makes the presence excess absorption near $3.45$~$\mu$m in our evening transmission spectrum intriguing. Recent modelling work by \cite{Ohno_2024} has also shown that hydrogenated diamond haze produces a distinct peak at 3.53~$\mu$m, however a detailed exploration of the existence of these more complex carbon-bearing molecules on the evening limb of \planet is beyond the scope of this study.
\subsection{Atmospheric limb retrievals}
\subsubsection{Free-chemistry retrievals}
We used the \texttt{Exo Skryer} atmospheric modelling and retrieval code to perform 1.5D atmospheric retrievals on our morning/evening spectra simultaneously (see Section~\ref{sec:exo_skryer_retrievals} for details). We first performed free-chemistry retrievals  with two model setups. The first allowed the temperatures to vary independently between the limbs, whereas the second run linked the morning and evening temperatures via Equation\,(\ref{eqn:linked_temps}). We find decisive evidence (${\Delta\ln Z=4.85\pm0.52}$; \citealt{Jeffreys_1935}) for the model which enforces $T_{\rm iso}^{\rm Morn}\le T_{\rm iso}^{\rm Eve}$. We find that if we don't enforce the morning temperature to not exceed the evening temperature via Equation\,(\ref{eqn:linked_temps}), the independent-limb temperature model yields a higher morning temperature ({$T^{\rm Morn}_{\rm iso}=1946_{-101}^{+121}~$K$~>T^{\rm Eve}_{\rm iso}=1816_{-111}^{+128}$~K}). This is a well-known artifact of the scale-height degeneracy \citep{benneke2013how-fd8,line2016influence-47e,heng2017theory-972,Welbanks_2019_On_Degeneracies_in_Retrievals_of_Exoplanetary_Transmission_Spectra}, where the retrieved temperature can compensate for changes in molecular abundances, mean molecular weight, and reference radius, among others, through their combined influence on the atmospheric scale height and resulting transmission spectrum. Therefore, we adopt the linked-temperature model as our fiducial free-chemistry retrieval as we do not expect the morning limb to be hotter than the evening limb due to eastward heat advection by an equatorial superrotating jet \citep{Showman_2009,kataria2016atmospheric-0be}. We also find a strong preference (${\Delta\ln Z=2.81\pm0.49}$) for our 1.5D models versus a symmetric model in which all parameters are shared between the limbs. The results of our fiducial 1.5D free-chemistry retrieval are shown in Figure~\ref{fig:freechem_retrieval}.

We find that \ce{H2O} and \ce{CO2} exhibit consistent retrieved abundances between the two limbs (see Figure~\ref{fig:freechem_retrieval} and Table~\ref{tab:free_chem}), disfavouring strong chemical asymmetries in these species. However, the \ce{H2O} abundances derived from our 1.5D retrieval are lower than our 1D retrieval and the constraints from the NIRCam-only retrieval presented by \cite{xue2024jwst-bb8}. These lower \ce{H2O} abundances ($\log_{10}\ce{H2O}^{\rm Eve} = -3.98_{-0.28}^{+0.30}$~and~$\log_{10}\ce{H2O}^{\rm Morn} = -4.09_{-0.32}^{+0.27}$) are consistent with previous retrievals across a broader wavelength range, in which they also included HST~STIS+WFC3 data \citep{Verma_2025}, as well as JWST~MIRI/LRS observations \citep{chubb2026magnesiumsilicatecloudsatmosphere}. 
In contrast, \ce{CH4} shows a significant abundance asymmetry between the two limbs, with the morning limb favouring a higher abundance ($\log_{10}\ce{CH4}^{\rm Morn} = -8.67_{-1.32}^{+1.23}$) than the evening limb ($\log_{10}\ce{CH4}^{\rm Eve} = -11.30_{-0.50}^{+0.68}$). \ce{CO} shows a similar, although weaker, trend ($\log_{10}\ce{CO}^{\rm Morn} = -7.27_{-1.43}^{+1.39}$ and $\log_{10}\ce{CO}^{\rm Eve} = -9.73_{-1.47}^{+1.99}$), with the evening posterior again extending toward the prior bound. It is worth noting that neither \ce{CO} nor \ce{CH4} were confidently detected in the original 1D retrieval analysis of this dataset by \citet{xue2024jwst-bb8}.

The thermal profiles show a marginally hotter evening limb (${T^{\rm Eve}_{\rm iso}=1897_{-104}^{+106}}$~K) with a small temperature offset of $\Delta T_{\rm iso}^{\rm Morn} = 45_{-29}^{+40}$~K. The cloud structures, however, show a more pronounced asymmetry. The evening cloud forms at higher altitude, with a retrieved cloud-base pressure of $\log_{10}(P_{\rm base}^{\rm Eve}/\mathrm{bar}) = -3.86_{-0.27}^{+0.30}$, compared to $\log_{10}(P_{\rm base}^{\rm Morn}/\mathrm{bar}) = -2.51_{-0.66}^{+0.83}$  for the morning limb, with minimal overlap between the two posteriors. The retrieved cloud base pressures are broadly consistent with previous studies that infer high-altitude cloud decks at pressures of $\sim0.01$--$0.1$ mbar \citep{macdonald2017hd-e41,fairman2024importanceopticalwavelengthdata}. The evening cloud also falls off more rapidly with increasing altitude above its base, as reflected in its steeper retrieved slope parameter ($\log_{10}\alpha_{\rm cld}^{\rm Eve} = 1.29_{-0.45}^{+0.42}$ vs. \ $\log_{10}\alpha_{\rm cld}^{\rm Morn} = 0.65_{-0.62}^{+0.56}$). Physically, this describes an evening cloud that is thin and vertically confined near its base, in contrast to a deeper, more vertically extended morning cloud. The retrieved temperature and cloud properties for the fiducial 1.5D free-chemistry retrieval are summarised in Table~\ref{tab:temp_cloud_comp}. The difference in cloud properties between the morning and evening limbs is also consistent with the fractional cloud coverage inferred by previous retrieval studies \citep[e.g.,][]{Barstow_2017,macdonald2017hd-e41,fairman2024importanceopticalwavelengthdata,xue2024jwst-bb8}, as well as our 1D retrieval (Table~\ref{tab:prt_priors}). These results show that the asymmetric model is statistically favoured over a symmetric model despite the additional free parameters, driven primarily by the cloud properties, which show the clearest morning--evening separation of any retrieved parameter. The gas chemistry and thermal structure exhibit comparatively weaker asymmetries, with \ce{H2O} and \ce{CO2} abundances and the isothermal temperatures remaining consistent between limbs within their marginal posteriors. This suggests that, despite \planet being a highly favourable target for limb asymmetry studies, large chemical and thermal contrasts are not ubiquitous and may require more extreme atmospheric conditions to become detectable at these wavelengths. The limited wavelength coverage and lack of sensitivity to continuum opacity between $3\text{--}5~\mu\text{m}$ may also limit our ability to robustly constrain cloud asymmetries, which have been shown to cause prominent variations between morning and evening transmission spectra at shorter wavelengths \citep{mukherjee2025cloudymorningsclearevenings,Murphy_2025}.

Although modest, the sign of this temperature asymmetry is consistent with three-dimensional GCMs, which predict that eastward equatorial jets transport hotter dayside material across the evening limb, while cooler nightside gas is transported across the morning terminator \citep[e.g.,][]{showman2002atmospheric-f14,parmentier2016transitions-9d2}. The relatively small magnitude of the retrieved temperature contrast suggests efficient heat redistribution across the terminators at the pressures probed by our observations. The lack of significant differences in \ce{H2O} and \ce{CO2} abundances between the morning and evening limbs also suggests that atmospheric transport is sufficient to maintain a relatively homogeneous terminator composition for these species. The detected asymmetry in the retrieved cloud properties is also consistent with theoretical predictions, in which clouds are expected to be enhanced on the morning limb as cooler nightside material is transported across the terminator \citep{parmentier2016transitions-9d2}.

\begin{figure*}[ht!]
    \centering
    \includegraphics[width=\linewidth]{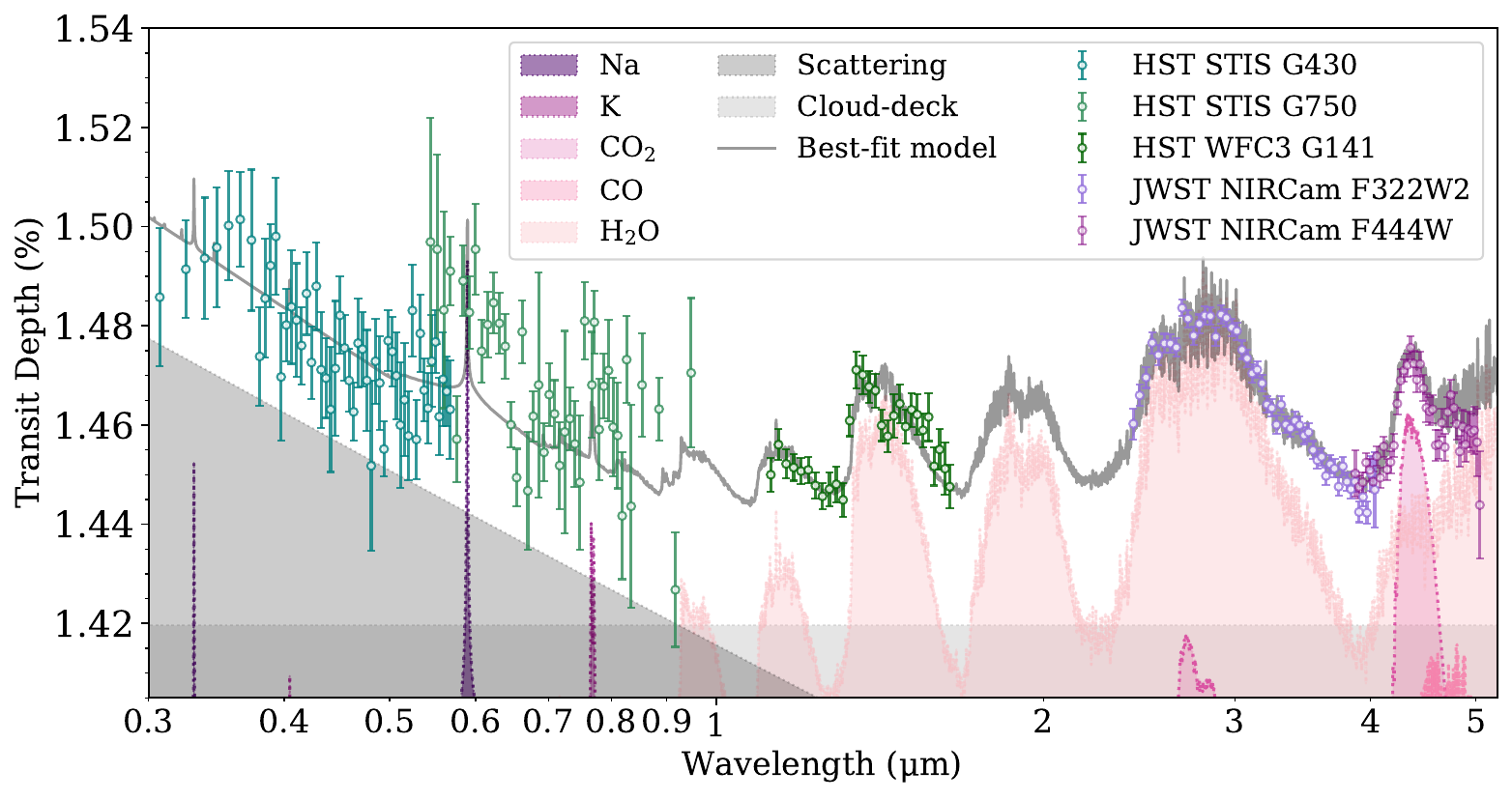}
    \caption{Transmission spectrum of \planet with the best-fit model (gray line) and opacity contributions (coloured) for the 1D atmospheric retrieval using \texttt{petitRADTRANS}. The green data points are the HST~STIS+WFC3 measurements from \cite{sing2016continuum-9d7} while the purple ones are the two transits observed by JWST~NIRCam using our uniform (limb-averaged) \texttt{Eureka!} reduction.}
    \label{fig:prt-1d-contributions}
\end{figure*}

\begin{deluxetable*}{lcc|cc}[htbp!]
\tablecaption{Posterior values of the temperature and cloud parameters from the 1.5D \texttt{Exo Skryer} retrievals of \planet.}
\label{tab:temp_cloud_comp}
\tablehead{
\colhead{} & \multicolumn{2}{c|}{Free-chemistry} & \multicolumn{2}{c}{Equilibrium Chemistry} 
}
\startdata
 & Morning & Evening & Morning & Evening \\[4pt]
$T_{\rm iso}$ [K]   & $1847^{+101}_{-100}$ & $1897_{-104}^{+106}$ & $1689_{-75}^{+93}$ & $1731_{-81}^{+93}$  \\
$\log_{10}P_{\rm base}$ [bar] & $-2.51_{-0.66}^{+0.83}$ & $-3.86_{-0.27}^{+0.30}$ & $-2.21_{-1.62}^{+2.40}$ & $-3.07_{-1.47}^{+2.50}$  \\
$\log_{10}\alpha_{\rm cld}$ & $0.65_{-0.62}^{+0.56}$ & $1.29_{-0.45}^{+0.42}$ & $0.81_{-0.77}^{+0.55}$ & $1.02_{-0.81}^{+0.55}$\\
$\log_{10}\kappa_{\rm cld}$ [cm$^{2}$\,g$^{-1}$] & $-0.50_{-0.27}^{+0.22}$ & $-0.24_{-0.28}^{+0.24}$ & $0.40_{-0.24}^{+0.23}$ & $0.40_{-0.19}^{+0.16}$\\[2pt]
\enddata
\end{deluxetable*}

\begin{deluxetable*}{lccc|cc}[htbp!]
% \tabletypesize{\normalsize} 
\tablecaption{Posterior values of the carbon-to-oxygen (C/O) ratio and metallicity ([M/H]) from the chemical equilibrium retrievals of \planet for various retrieval setups and wavelength coverages. ES and pRT denote \texttt{Exo Skryer} and \texttt{petitRADTRANS} retrievals, respectively.}
\label{tab:chem_eq_comp}
\tablehead{
\colhead{} & \multicolumn{3}{c|}{1.5D Retrieval (NIRCam)} & \multicolumn{2}{c}{1D Retrieval} 
}
\startdata
\multicolumn{1}{c}{} & \multicolumn{1}{c}{Morning (ES)} & \multicolumn{1}{c}{Evening (ES)} & \multicolumn{1}{c|}{Shared Chem. (ES)} & \multicolumn{1}{c}{NIRCam (ES)} & \multicolumn{1}{c}{NIRCam+HST (pRT)} \\[4pt] 
$\rm [M/H]$ & $0.34_{-0.35}^{+0.25}$ & $0.57_{-0.23}^{+0.26}$ & $0.68_{-0.24}^{+0.15}$ & $0.66_{-0.14}^{+0.16}$ & $0.24\pm0.14$ \\
$\rm C/O$   & $0.26_{-0.10}^{+0.18}$ & $0.28_{-0.12}^{+0.18}$ & $0.39_{-0.13}^{+0.14}$ & $0.12_{-0.02}^{+0.04}$ & $0.14_{-0.03}^{+0.06}$ \\[2pt]
\enddata
\end{deluxetable*}
\clearpage
For our 1.5D retrievals, both free and equilibrium chemistry, we consistently find higher inferred temperatures on our limbs when compared to that inferred from our 1D retrieval (see Figures~\ref{fig:freechem_retrieval}~\&~\ref{fig:equilchem_retrieval}), and the photospheric temperatures inferred from previous 1D atmospheric retrieval studies of \planet \citep{xue2024jwst-bb8,Verma_2025,Bachmann_2025,chubb2026magnesiumsilicatecloudsatmosphere}. This is consistent with the findings of \cite{macdonald2020why-0e8}, in which they demonstrate that 1D retrievals of transmission spectra erroneously infer cooler temperature profiles when applied to datasets in which multidimensional effects are present. This erroneous inference of a cooler temperature by our 1D retrieval may also cause an artificial increase in the \ce{H2O} abundance relative to our 1.5D retrievals, as highlighted above, in order to match the extent of the \ce{H2O} feature.

\subsubsection{Equilibrium chemistry retrievals}
We also performed equilibrium chemistry retrievals on our morning/evening spectra with two distinct model setups. The first shared both C/O and [M/H] between the limbs, whereas the second allowed C/O and [M/H] to vary independently on each limb. For both modelling setups, the retrieved C/O and [M/H] values converge to similar constraints along the terminator (see Figure~\ref{fig:equilchem_retrieval}). The morning and evening limbs both retrieve super-solar metallicities 
%(${\text{[M/H]}_{\rm Morn}=0.98_{-0.19}^{+0.16}}$ and ${\text{[M/H]}_{\rm Eve}=1.06_{-0.14}^{+0.11}}$), 
and sub-solar carbon-to-oxygen ratios (see Table~\ref{tab:chem_eq_comp}). 
%(${\text{C/O}_{\rm Morn}=0.40_{-0.16}^{+0.15}}$ and ${\text{C/O}_{\rm Eve}=0.39_{-0.13}^{+0.14}}$). 
These independent limb constraints, as well as our shared chemistry retrieval, agree closely with both the stellar C/O (${\text{C/O}=0.42\pm0.08}$, \citealt{polanski2022chemical-d17}) and the combined NIRCam$+$WFC3 equilibrium constraints from \citet{xue2024jwst-bb8}. This chemical homogeneity across the limbs reinforces the conclusion drawn from our free-chemistry retrieval, in which the atmosphere of \planet does not exhibit a chemical asymmetry for the most prominent species investigated across the observed wavelength range. The shared chemistry C/O constraints obtained from our \texttt{Exo Skryer} 1.5D retrieval are also consistent with those derived from the independent 1.5D \texttt{petitRADTRANS} retrieval (see Table~\ref{tab:prt_priors}). As the limb-independent chemical equilibrium retrieval finds a uniform composition, the variation in scale height seen in Figure~\ref{fig:catharm_eureka} between the morning and evening limbs is driven primarily by slight variations in the thermal properties. We retrieve an evening limb that is marginally hotter ({$T^{\rm Eve}_{\rm iso}=1731_{-81}^{+93}$~K}), contrasted against a slightly cooler ({$T^{\rm Morn}_{\rm iso}=1689_{-75}^{+93}$~K}) morning limb, corresponding to a temperature contrast of approximately 40 K. We also find no significant variation in cloud properties, in contrast to our 1.5D free-chemistry retrieval. The retrieved temperature and cloud properties for the 1.5D free and equilibrium retrievals are summarised in Table~\ref{tab:temp_cloud_comp}.

% This is in line with theoretical predictions from GCMs, in which clouds preferentially condense on the cooler nightside before being transported to the morning limb, where they persist at lower pressures compared to the evening limb \citep{helling2016mineral-c35,Chubb_2024}.

In contrast to our 1.5D retrievals, our 1D NIRCam-only equilibrium chemistry retrieval finds a highly sub-solar carbon-to-oxygen ratio (${\rm C/O = 0.12_{-0.02}^{+0.04}}$) and a super-solar metallicity (${{\rm [M/H] = 0.67 \pm 0.15}}$; see Table~\ref{tab:chem_eq_comp} \& Figure~\ref{fig:equilchem_retrieval}). While these C/O constraints are in agreement with the 1D NIRCam-only constraints presented by \citet{xue2024jwst-bb8}, they diverge significantly from the findings of our 1.5D retrievals. This is in alignment with the decrease in the retrieved \ce{H2O} abundance seen as we go from 1D to 1.5D retrievals, while the other chemical abundances remain consistent (see Figure~\ref{fig:freechem_retrieval}). 
% We note that this consistency across different modelling assumptions and literature values is largely driven by the broad nature of the retrieved posterior distributions of our 1.5D retrievals in Figure~\ref{fig:equilchem_retrieval}.
When we include optical data in our \texttt{petitRADTRANS} 1D NIRCam+HST equilibrium chemistry retrieval (see Table~\ref{tab:chem_eq_comp} \& Figure~\ref{fig:prt-1d-contributions}), we now find only a slightly super-solar metallicity and a sub-solar C/O ratio. Recent studies have also incorporated both optical HST~STIS \citep{Verma_2025} and mid-infrared JWST~MIRI/LRS \citep{chubb2026magnesiumsilicatecloudsatmosphere} data in their 1D retrievals alongside the JWST~NIRCam data reanalysed as part of this study. The inclusion of optical data has resulted in lower inferred H$_2$O and CO$_2$ abundances, which introduces a shift in the derived C/O and [M/H] values compared to analysing near-infrared data alone. Both studies retrieve significantly lower metallicities ($\text{[M/H]} \approx 0.1\text{--}0.2$) compared to near-infrared only analyses, consistent with our findings. However, \citet{Verma_2025} retrieve a solar C/O ratio (${\text{C/O}=0.56_{-0.12}^{+0.10}}$), whereas \citet{chubb2026magnesiumsilicatecloudsatmosphere} analysed the full $0.6\text{--}12~\mu\text{m}$ transmission spectrum of \planet, and directly detected magnesium silicate cloud absorption whilst retrieving a sub-solar carbon-to-oxygen ratio (${\text{C/O}=0.16_{-0.04}^{+0.06}}$). The inclusion of both optical and mid-infrared data allows for a continuum cloud opacity which truncates the near-infrared molecular features, and aligns the C/O with both previous literature values as well as the {1D~NIRCam+HST} C/O constraints presented in this work. Figure~\ref{fig:co_mh_litcomp} shows a comprehensive comparison of retrieved [M/H] and C/O from this study as well as previous literature values. As the JWST~MIRI/LRS observations only captured a partial transit of \planet, and consequently missed the transit egress, they are not conducive to limb asymmetry analysis and thus were not analysed as part of this study.

In each of our 1.5D atmospheric retrievals, we consistently find isothermal temperatures which are larger than {$T_{\rm eq}=1450~$K}. This is due to the inherent limitation of assuming a vertically isothermal temperature profile when the true atmosphere likely possesses an altitude-dependent thermal gradient. The isothermal temperature profile acts as a pressure-weighted average of the true temperature gradient along the slant path. Because transmission spectroscopy is primarily sensitive to the low-pressure, high-altitude limbs, the model is biased toward the temperatures of these upper layers. Secondly, transmission observations observe the terminator, which is subject to stellar irradiation and day-night transport. It has been shown that 1D $T$--$P$ profiles cannot resolve this longitudinal day-night temperature gradient, biasing inferred temperatures to higher temperatures than the true terminator temperature \citep{Caldas_2019}. Finally, we are also subject to the aforementioned scale-height degeneracy, where the retrieval can force an artificially high temperatures to compensate for aerosol-induced spectral slopes or variations in absolute molecular abundances that are unknown due to our narrow wavelength range.

While our equilibrium chemistry retrievals indicate a terminator region which has a cooler morning limb and a marginally hotter evening limb, in agreement with theoretical predictions, we caution against the over-interpretation of inferred chemical, temperature, or cloud properties derived from atmospheric retrievals over narrow wavelength ranges. Our morning and evening spectra are particularly limited by a lack of a sufficient continuum opacity between $3\text{--}5~\mu\text{m}$. As shown by previous works at this wavelength range \citep[e.g.,][]{fairman2026bowie-align-131}, this limitation, compounded in our study by transit depth uncertainties that are significantly larger than those of a uniform light curve model, allows for a wide range of degenerate solutions, making it difficult to ascertain constraints on the variation of atmospheric parameters between the limbs. 

\section{Conclusions} \label{sec:conclusions}
In this study, we presented the first limb asymmetry analysis of the canonical hot Jupiter \planet, a compelling target for limb asymmetry due to its brightness, large atmospheric scale height and temperature. We independently reduced and reanalysed archival JWST~NIRCam~F322W2 \& F444W transit observations first presented in \citealt{xue2024jwst-bb8} to conduct our study. Our findings are summarised below:
\begin{itemize}
    \item In order to ensure any observed limb asymmetries arise due to atmospheric inhomogeneity, as opposed to uncertainties in transit timing or orbital configuration, we refined the HD~209458 system parameters via a simultaneous fit to transit and RV data, spanning more than 25 years, using the \texttt{juliet} fitting package. This fit improved the precision on the system parameters by a factor of $\sim$2--3 relative to previous literature values.
    
    \item We utilised two light curve fitting routines, \texttt{Harmonica} and \texttt{catwoman}, to search for limb asymmetries in transit. We find the extracted morning and evening transmission spectra to be in excellent agreement between both fitting routines.

    \item We performed 1.5D atmospheric retrievals on our extracted limb spectra, using the \texttt{Exo Skryer} retrieval code. Our free-chemistry retrievals strongly prefer an asymmetric atmosphere over a symmetric atmosphere, with the clearest morning--evening asymmetry found in the retrieved cloud properties, and a weaker asymmetry in the \ce{CH4} and \ce{CO} abundances. In contrast, the \ce{H2O} and \ce{CO2} abundances and the isothermal temperature structure are largely consistent between limbs. These results demonstrate that even for a highly favourable target, strong chemical and thermal limb asymmetries may not be readily detectable, particularly over a narrow wavelength range with limited continuum opacity. We also find limb temperatures, \ce{H2O} abundances, and cloud opacity constraints which diverge significantly from those obtained from our 1D retrievals.
    
    \item Our 1.5D equilibrium chemistry retrievals also find an atmosphere which has uniform chemical composition across the limbs, however we find an evening limb which is marginally hotter, compared to a cooler morning limb. We find a sub-solar C/O ratio from our 1.5D retrievals for both the morning and evening limbs, aligning with the host star C/O. We also find a super-solar metallicity, consistent with our 1D retrieval and previous works across this wavelength range. Similarly to our 1.5D free-chemistry retrievals, we find higher temperatures and cloud opacities, as well as differing C/O, relative to our 1D retrievals. These results highlight the importance of considering multidimensional effects in order to mitigate known 1D retrieval biases.

    \item We also perform 1D retrievals using the \texttt{petitRADTRANS} retrieval code, including HST~STIS \& WFC3 observations alongside the NIRCam data. We find the inclusion of optical data shifts the retrieved metallicity towards lower, slightly super-solar values, consistent with previous optical and infrared retrieval studies of \planet \citep{Verma_2025,chubb2026magnesiumsilicatecloudsatmosphere}. We retrieve sub-solar carbon-to-oxygen ratios, consistent with our 1.5D retrievals, as well as with previous retrieval studies of \planet \citep{xue2024jwst-bb8,Bachmann_2025,chubb2026magnesiumsilicatecloudsatmosphere}. 
    
\end{itemize}

%% Please use the acknowledgment and contribution environments. This will 
%% be anonomyized when the "anonymous" style option is used. 
\begin{acknowledgments}
This work is based on observations made with the NASA/ESA/CSA James Webb Space Telescope. The data were obtained from the Mikulski Archive for Space Telescopes at the Space Telescope Science Institute, which is operated by the Association of Universities for Research in Astronomy, Inc., under NASA contract NAS 5-03127 for JWST. These observations are associated with program \#1274.

CM and HRW were funded by UK Research and Innovation (UKRI) framework under the UK government’s Horizon Europe funding guarantee for an ERC Starter Grant [grant number EP/Y006313/1]. CM thanks Siddharth Gandhi for a helpful discussion on reparameterising 1.5D retrievals.
% I asked Everett whether there's any updates to the NIRCam JWST pipeline / calibration files that could make our uncertainties smaller. 
% thank Ryan for help unless he'll become a co-author
% 

\end{acknowledgments}

% \begin{contribution}
% %%This section gives authors the space to recognize author contributions. The text inside this environment is NOT counted towards the total word quanta. At a minimum, manuscripts are expected to include this text:

% All authors contributed equally to the Terra Mater collaboration.

% %% But authors are expected to provide more specific details, e.g. 
% %%
% %%SC was responsible for writing and submitting the manuscript.
% %%WWM came up with the initial research concept and edited the manuscript.
% %%OTS obtained the funding and edited the manuscript.
% %%EBF provided the formal analysis and validation. He also edited the manuscript.
% %%GEH Supervised the undergraduates, wrote the software and administers the project github and Zenodo repositories.
% %%
% %% Authors can use the Contributor Role Taxonomy (CRediT) at
% %% https://credit.niso.org
% %% for ideas on how write a good statement tailored to their needs.

% \end{contribution}

\facilities{CORALIE \citep{Coralie_queloz_2000}, ELODIE \citep{Elodie_1996}, ESPRESSO \citep{espresso_pepe_2021}, HARPS \citep{mayor_harps_2003}, HIRES \citep{Vogt_hires_1994}, HST~STIS \citep{woodgate_stis_1997,kimble_stis_1998}, HST~WFC3 \citep{Mackenty_wfc3_2010}, JWST~NIRCam \citep{rieke2005overview-0e9,Rieke_nircam_2023}, TESS \citep{ricker2015transiting-029}.}

\software{NumPy \citep{Numpy}, SciPy \citep{Scipy}, astropy \citep{astropy}, iPython \citep{ipython}, Matplotlib \citep{matplotlib}, Eureka! \citep{bell2022eureka-6ec}, Tiberius \citep{kirk2017rayleigh-ae0,kirk2021access-85d}, juliet \citep{espinoza_2019_juliet}, corner \citep{corner}, Exo Skryer \citep{Lee_2026_exoskryer}, petitRADTRANS \citep{mollire2019petitradtrans-7c8, nasedkin2024atmospheric-ed7, blain2024spectralmodel-d73}.}

\appendix
\label{sec:appendix}
\setcounter{figure}{0}
\renewcommand{\thefigure}{A\arabic{figure}}
\setcounter{table}{0}
\renewcommand{\thetable}{A\arabic{table}}
\renewcommand{\theHfigure}{app.\arabic{figure}}
\renewcommand{\theHtable}{app.\arabic{table}}

\begin{figure*}[ht!]
\includegraphics[width=\linewidth]{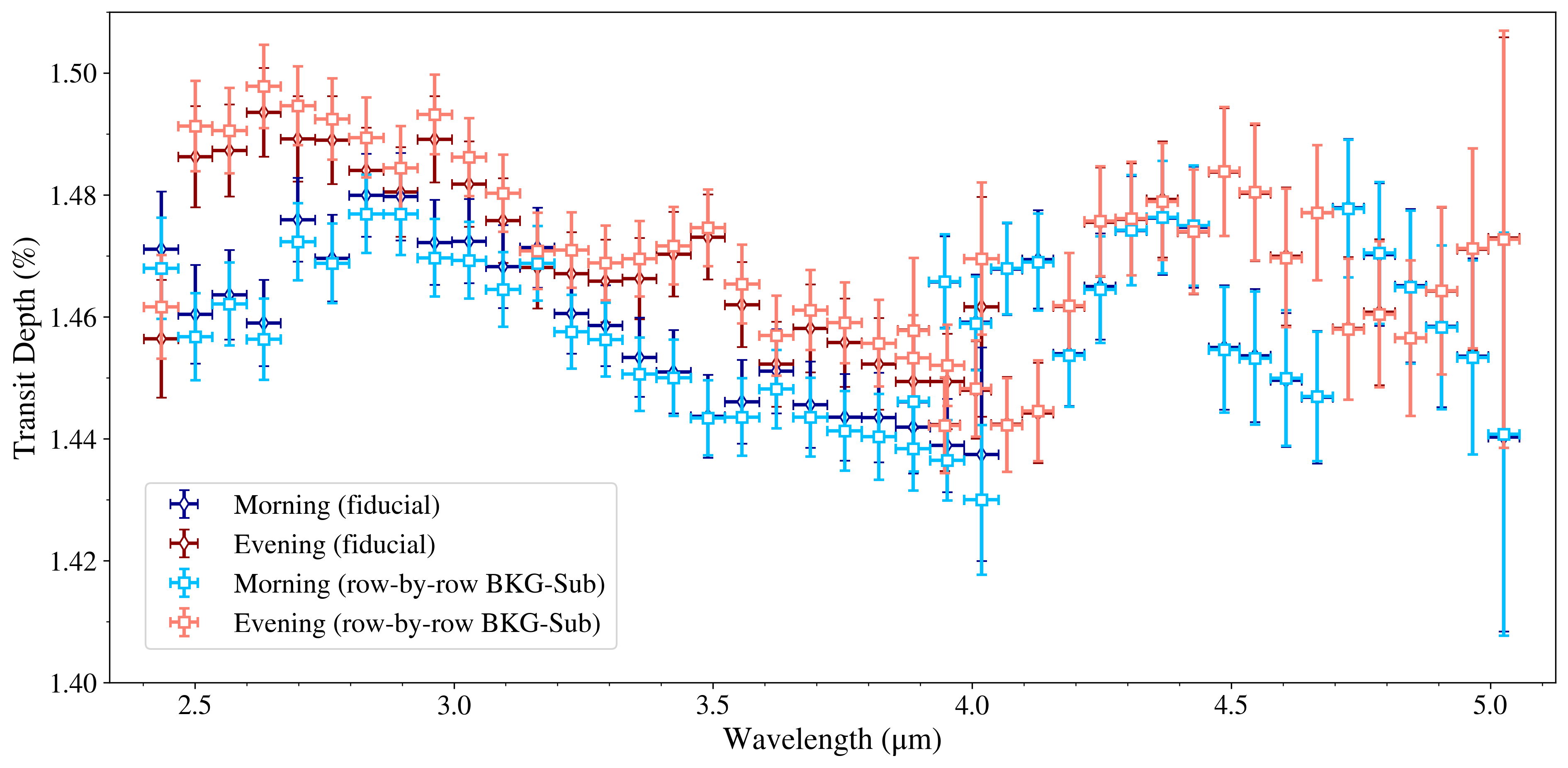}
\caption{Morning and evening transmission spectra obtained from \texttt{Harmonica} light curve fitting to two different \texttt{Eureka!} reductions, outlined in Section~\ref{sec:eureka}. The white diamonds show the spectra obtained from our fiducial reduction, whereas the white squares include a group-level row-by-row background subtraction step to mitigate potential 1/f noise.}
\label{fig:rowbyrow_spectra}
\end{figure*}
\begin{deluxetable}{lccc}[ht!]
\tablewidth{\columnwidth}
\tablecaption{Prior and posterior values for the orbital parameters from the \texttt{juliet} fit of \planet assuming an eccentric orbit.}
\label{tab:transit_rv_ecc}
\tablehead{
\colhead{Parameter} & \colhead{Prior} & \colhead{Posterior} & \colhead{Units} 
}
\startdata
\multicolumn{1}{l}{\textit{Transit parameters}}\\
$P$ & $\mathcal{N}(3.52474859,10^{-6})$ & $3.52474893 \pm 0.00000014$ & days \\
$t_0$ (BJD$_{\rm TDB}-2450000$) & $\mathcal{N}(9893.75125,10^{-4})$ & $9893.751205 \pm 0.000010$ & days \\
$a/R_\star$ & $\mathcal{U}(8.5,9.0)$ & $8.718 \pm 0.029$ & --- \\
$b$ & $\mathcal{U}(0,1)$ & $0.5101 \pm 0.0018$ & --- \\
$e$ & $\mathcal{N}_{[0,1]}(0.01,0.011)$ & $0.0112 \pm 0.0028$ & --- \\
$\omega$ & $\mathcal{U}(0,360)$ & $62.1^{+10.7}_{-7.5}$ & deg \\
$R_{\rm{p}}^{\rm TESS~S56~(SC)}$ & $\mathcal{U}(0.1195,0.122)$ & $0.12103 \pm 0.00005$ & $R_\star$ \\
$R_{\rm{p}}^{\rm TESS~S82~(SC)}$ & $\mathcal{U}(0.1195,0.122)$ & $0.12106 \pm 0.00006$ & $R_\star$ \\
$R_{\rm{p}}^{\rm NIRCam~F322W2}$ & $\mathcal{U}(0.1195,0.122)$ & $0.12105 \pm 0.00004$ & $R_\star$ \\
$R_{\rm{p}}^{\rm NIRCam~F322W2~(SW)}$ & $\mathcal{U}(0.1195,0.122)$ & $0.12126 \pm 0.00011$ & $R_\star$ \\
$R_{\rm{p}}^{\rm NIRCam~F444W}$ & $\mathcal{U}(0.1195,0.122)$ & $0.12082 \pm 0.00005$ & $R_\star$ \\
$R_{\rm{p}}^{\rm NIRCam~F444W~(SW)}$ & $\mathcal{U}(0.1195,0.122)$ & $0.12030 \pm 0.00020$ & $R_\star$ \\ \\
\multicolumn{1}{l}{\textit{RV parameters}} \\
$K_{\rm p}$ & $\mathcal{N}(83.5,5)$ & $83.4 \pm 0.4$ & $\rm m~s^{-1}$ \\
$m$ & $\mathcal{U}(-10^{-2},10^{-2})$ & $-0.0002 \pm 0.0004$ & $\rm m~s^{-1}~day^{-1}$ \\
$c$ & 0 (fixed) & --- & $\rm m~s^{-1}$ \\
\enddata
\tablenotetext{}{$\mathcal{N}_{[a,b]}(\mu, \sigma)$ denotes a normal distribution of mean $\mu$ and standard deviation $\sigma$, truncated within the bounds $[a, b]$.}
\end{deluxetable}
\begin{figure*}[ht!]
\includegraphics[width=\linewidth]{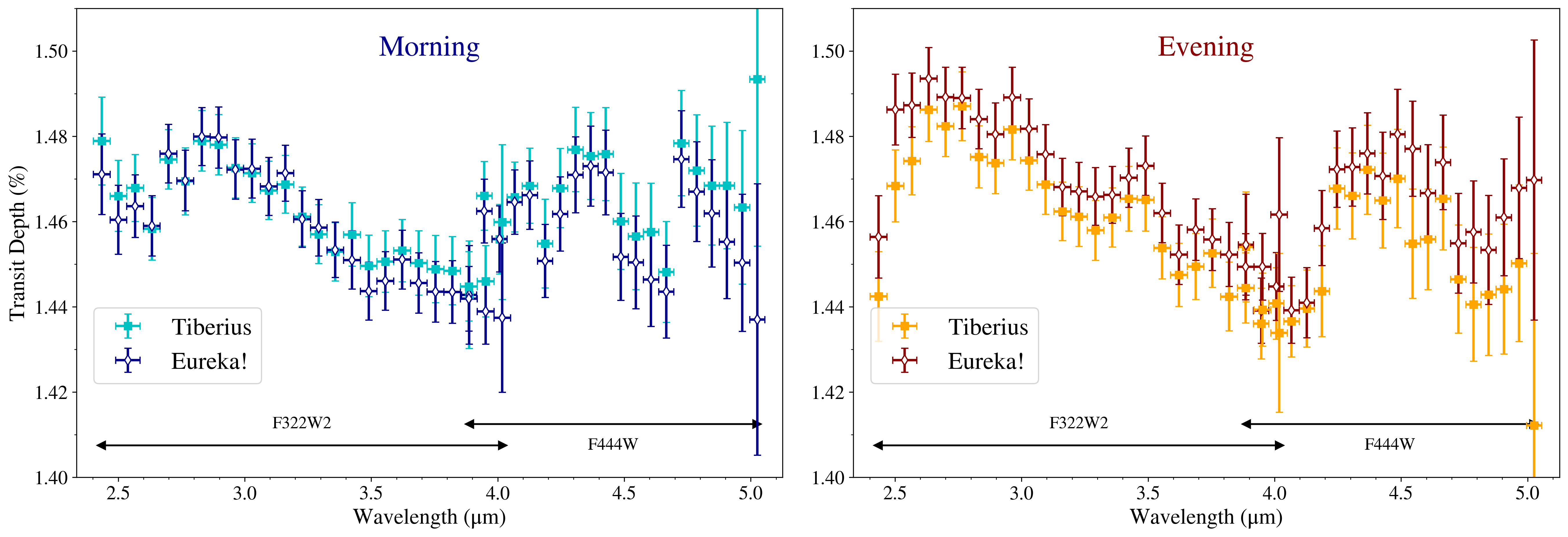}
\caption{The morning (left) and evening (right) transmission spectra obtained for each of our reduction pipelines (detailed in Sections~\ref{sec:eureka}~\&~\ref{sec:tiberius}). The light curves were fit with \texttt{Harmonica} (see Section~\ref{sect:harmonica_fitting}). The transit depths obtained from the \texttt{Tiberius} reduction are shown as filled squares, whereas the transit depths obtained from the \texttt{Eureka!} reduction are shown as unfilled diamonds. Horizontal black arrows highlight the wavelength coverage of the NIRCam~F322W2 and F444W filters. No offset has been applied to the F444W values.}
\label{fig:catharm_tiberius}
\end{figure*}
\begin{figure*}[ht!]
\includegraphics[width=\linewidth]{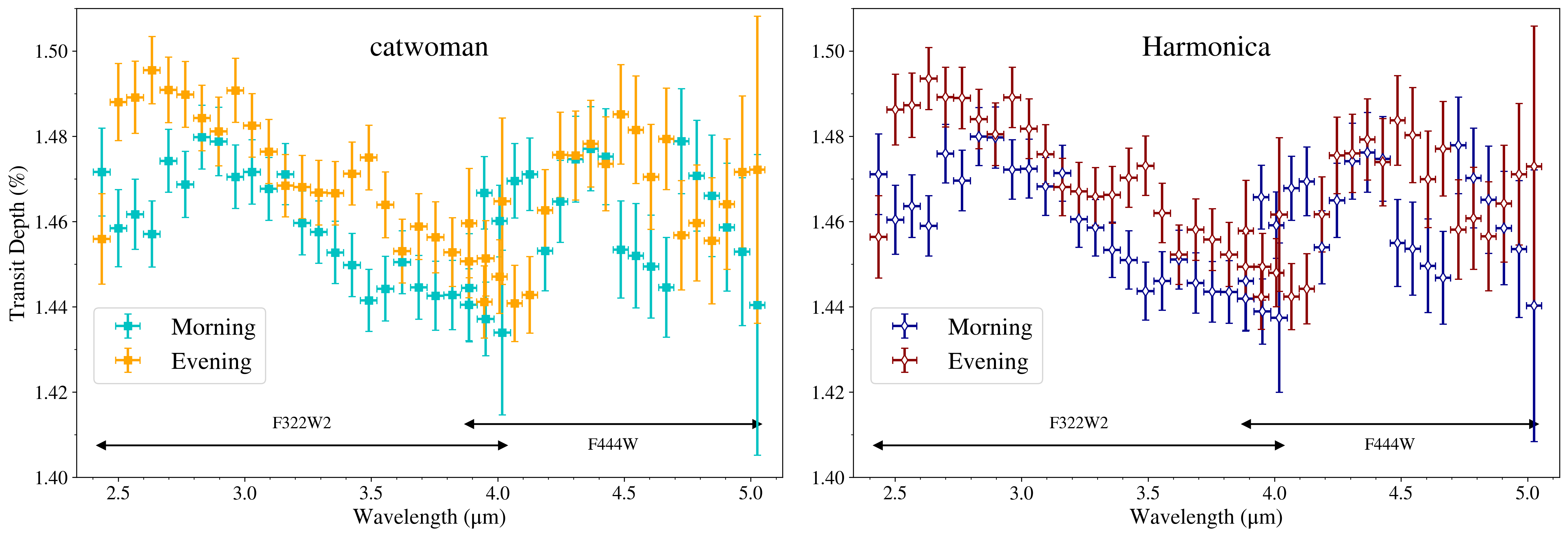}
\caption{(left) A comparison of the morning (cyan) and evening (orange) transmission spectra obtained from our \texttt{catwoman} fit (see Section~\ref{sec:catwoman_fitting}). (right) A comparison of the morning (blue) and evening (red) transmission spectra obtained from our \texttt{Harmonica} fit (see Section~\ref{sect:harmonica_fitting}). The spectra shown here were reduced with \texttt{Eureka!}. The best-fit offset value from our fiducial 1.5D free-chemistry atmospheric retrieval has been applied to the F444W values for clarity.}
\label{fig:morneve_comp}
\end{figure*}
% \begin{figure*}[ht!]
% \includegraphics[width=\linewidth]{HD209_pRT_EquilChemRetrieval_wCovar.png}
% \caption{The best-fit morning and evening transmission spectra from our 1.5D \texttt{petitRADTRANS} equilibrium chemistry retrievals, where the covariance between the morning and evening transit depths is accounted for in the likelihood.}
% \label{fig:pRT_cov}
% \end{figure*}
\begin{figure*}[ht!]
\includegraphics[width=\linewidth]{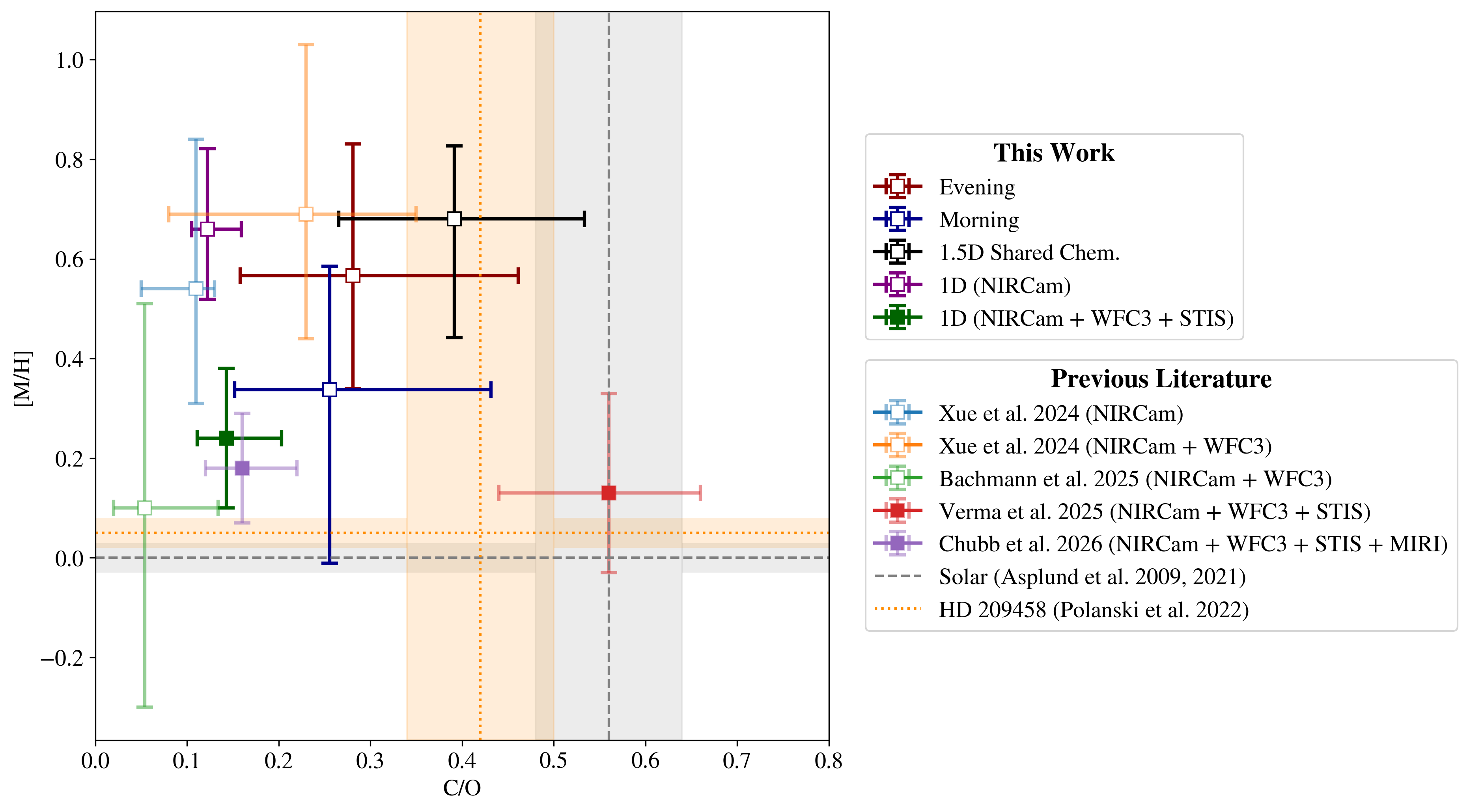}
\caption{The atmospheric metallicity, [M/H], and carbon-to-oxygen ratio, C/O, derived from various atmospheric retrievals of \planet. The opaque markers are the results presented in this study, whereas markers with a lower opacity are from previous literature values \citep[e.g.,][]{xue2024jwst-bb8,Bachmann_2025,Verma_2025,chubb2026magnesiumsilicatecloudsatmosphere}. Filled squares highlight studies which included optical data in their retrievals, which typically results in a lower retrieved [M/H] compared to analysing infrared data alone (unfilled squares). The vertical and horizontal grey dashed lines and shaded regions represent the solar values and their respective 1$\sigma$ uncertainties \citep{asplund2009chemical_comp,asplund_2021}. The vertical and horizontal orange dotted lines and shaded regions represent the stellar values and their respective 1$\sigma$ uncertainties \citep{polanski2022chemical-d17}. Symmetric uncertainties should be interpreted with caution, as they may be derived from non-Gaussian posterior distributions.} 
\label{fig:co_mh_litcomp}
\end{figure*}
\begin{table*}[t]
    \centering    
    \begin{minipage}[t]{0.48\textwidth}
        \centering
        \caption{Prior and posterior values for the instrumental parameters from the joint fit of \planet\ assuming a circular orbit. This includes flux offsets ($M$) and white-noise jitter terms ($\sigma_w$) for TESS and JWST, as well as RV offsets ($\mu$) and jitter ($\sigma_w$).}
        \label{tab:transit_rv_inst}
        \scriptsize
        \begin{tabular*}{\textwidth}{@{\extracolsep{\fill}}lccc}
            \toprule
            Parameter & Prior & Posterior & Units \\ 
            \hline
            \multicolumn{4}{@{}l}{\textit{Transit parameters}} \\
            $D^{\rm inst}$ & 1 (fixed) & --- & --- \\
            $M^{\rm TESS~S56~(SC)}$ & $\mathcal{N}(0,10^{2})$ & $3.8 \pm 6.1$ & ppm \\
            $M^{\rm TESS~S82~(SC)}$ & $\mathcal{N}(0,10^{2})$ & $-7.1 \pm 6.8$ & ppm \\
            $M^{\rm NIRCam~F322W2}$ & $\mathcal{N}(0,10^{2})$ & $-9.0 \pm 5.4$ & ppm \\
            $M^{\rm NIRCam~F322W2~(SW)}$ & $\mathcal{N}(0,10^{2})$ & $4.2 \pm 15.6$ & ppm \\
            $M^{\rm NIRCam~F444W}$ & $\mathcal{N}(0,10^{2})$ & $5.1 \pm 7.3$ & ppm \\
            $M^{\rm NIRCam~F444W~(SW)}$ & $\mathcal{N}(0,10^{2})$ & $-38.1 \pm 25.9$ & ppm \\
            $\sigma_w^{\rm TESS~S56~(SC)}$ & $\mathcal{J}(10,10^3)$ & $286.1 \pm 7.4$ & ppm \\
            $\sigma_w^{\rm TESS~S82~(SC)}$ & $\mathcal{J}(10,10^3)$ & $364.1 \pm 7.0$ & ppm \\
            $\sigma_w^{\rm NIRCam~F322W2}$ & $\mathcal{J}(10^{-6},10)$ & $< 0.9$ & ppm \\
            $\sigma_w^{\rm NIRCam~F322W2~(SW)}$ & $\mathcal{J}(10^{-6},10)$ & $< 0.7$ & ppm \\
            $\sigma_w^{\rm NIRCam~F444W}$ & $\mathcal{J}(10^{-6},10)$ & $< 0.7$ & ppm \\
            $\sigma_w^{\rm NIRCam~F444W~(SW)}$ & $\mathcal{J}(10^{-6},10)$ & $< 0.9$ & ppm \\ \\ 
            \multicolumn{4}{@{}l}{\textit{RV parameters}} \\
            $\mu^{\rm CORALIE}$ & $\mathcal{N}(0,10^{2})$ & $3.9 \pm 3.1$ & $\rm m~s^{-1}$ \\
            $\mu^{\rm ELODIE}$ & $\mathcal{N}(0,10^{2})$ & $0.2 \pm 3.0$ & $\rm m~s^{-1}$ \\
            $\mu^{\rm ESPRESSO}$ & $\mathcal{N}(0,10^{2})$ & $-1.25 \pm 0.19$ & $\rm m~s^{-1}$ \\
            $\mu^{\rm HARPS}$ & $\mathcal{N}(0,10^{2})$ & $-40.9 \pm 2.7$ & $\rm m~s^{-1}$ \\
            $\mu^{\rm HIRES~(pre~2004)}$ & $\mathcal{N}(0,10^{2})$ & $8.8 \pm 2.6$ & $\rm m~s^{-1}$ \\
            $\mu^{\rm HIRES~(post~2004)}$ & $\mathcal{N}(0,10^{2})$ & $5.5 \pm 1.5$ & $\rm m~s^{-1}$ \\
            $\sigma_w^{\rm CORALIE}$ & $\mathcal{J}(10^{-1},10^{2})$ & $12.8 \pm 1.3$ & $\rm m~s^{-1}$ \\
            $\sigma_w^{\rm ELODIE}$ & $\mathcal{J}(10^{-1},10^{2})$ & $7.1 \pm 2.1$ & $\rm m~s^{-1}$ \\
            $\sigma_w^{\rm ESPRESSO}$ & $\mathcal{J}(10^{-1},10^{2})$ & $1.47 \pm 0.13$ & $\rm m~s^{-1}$ \\
            $\sigma_w^{\rm HARPS}$ & $\mathcal{J}(10^{-1},10^{2})$ & $9.1 \pm 1.5$ & $\rm m~s^{-1}$ \\
            $\sigma_w^{\rm HIRES~(pre~2004)}$ & $\mathcal{J}(10^{-1},10^{2})$ & $5.5 \pm 0.6$ & $\rm m~s^{-1}$ \\
            $\sigma_w^{\rm HIRES~(post~2004)}$ & $\mathcal{J}(10^{-1},10^{2})$ & $4.7 \pm 0.6$ & $\rm m~s^{-1}$ \\
            \hline
            % \tablenotetext{}{Systemic velocity offsets were applied to bring all radial-velocity datasets to a common zero-point. The offsets and original datasets are available at \url{https://dace.unige.ch/radialVelocities/?pattern=hd\%20209458}.}
        \end{tabular*}
        \begin{flushleft}
        Systemic velocity offsets were applied to bring all radial-velocity datasets to a common zero-point. The offsets and original datasets are available at \url{https://dace.unige.ch/radialVelocities/?pattern=hd\%20209458}.
        \end{flushleft}
    \end{minipage}
    \hfill 
    \begin{minipage}[t]{0.48\textwidth}
        \centering
        % Ret72
        \caption{Prior and posterior values for the atmospheric parameters from the 1.5D \texttt{Exo Skryer} free-chemistry retrieval of \planet.}
        \label{tab:free_chem}
        \vspace{2mm}
        \scriptsize
        \begin{tabular*}{\textwidth}{@{\extracolsep{\fill}}lccc}
            \toprule
            Parameter & Prior & Posterior & Units \\ 
            \hline
            \multicolumn{4}{@{}l}{\textit{Shared}} \\
            $\log_{10}P_{\rm ref}$ & $\mathcal{U}(-2, 2)$ & $1.45_{-0.32}^{+0.31}$ & bar \\
            $\log_{10}g$ & $\mathcal{U}(2.5, 3.5)$ & $2.535_{-0.022}^{+0.026}$ & cm\,s$^{-2}$ \\
            ${\rm F444\rm W}_{\rm offset}$ & $\mathcal{U}(-200, 200)$ & $32.2_{-18.6}^{+19.0}$ & ppm \\ \\
            
            \multicolumn{4}{@{}l}{\textit{Morning}} \\
            $\Delta T_{\rm iso}$ & $\mathcal{U}(0, 500)$ & $45_{-29}^{+40}$ & K \\
            $\log_{10}\rm H_2O$ & $\mathcal{U}(-12, -2)$ & $-4.09_{-0.32}^{+0.27}$ & --- \\
            $\log_{10}\rm CO_2$ & $\mathcal{U}(-12, -2)$ & $-6.41_{-0.37}^{+0.34}$ & ---\\
            $\log_{10}\rm CO$ & $\mathcal{U}(-12, -2)$ & $-7.27_{-1.43}^{+1.39}$ & ---\\
            $\log_{10}\rm CH_4$ & $\mathcal{U}(-12, -2)$ & $-8.67_{-1.32}^{+1.23}$ & --- \\
            $\log_{10}P_{\rm base}$ & $\mathcal{U}(-6, 2)$ & $-2.51_{-0.66}^{+0.83}$ & bar \\
            $\log_{10}\alpha_{\rm cld}$ & $\mathcal{U}(-2, 2)$ & $0.65_{-0.62}^{+0.56}$ & --- \\
            $\log_{10}\kappa_{\rm cld}$ & $\mathcal{U}(-6, 6)$ & $-0.50_{-0.27}^{+0.22}$ & cm$^{2}$\,g$^{-1}$ \\
            $\log_{10}q_{\rm c}$ & $-1$ (fixed) & --- & g g$^{-1}$ \\ \\
            
            \multicolumn{4}{@{}l}{\textit{Evening}} \\
            $T_{\rm iso}$ & $\mathcal{U}(500, 3000)$ & $1897_{-104}^{+106}$ & K \\
            $\log_{10}\rm H_2O$ & $\mathcal{U}(-12, -2)$ & $-3.98_{-0.28}^{+0.30}$ & --- \\
            $\log_{10}\rm CO_2$ & $\mathcal{U}(-12, -2)$ & $-6.52_{-0.51}^{+0.42}$ & --- \\
            $\log_{10}\rm CO$ & $\mathcal{U}(-12, -2)$ & $-9.73_{-1.47}^{+1.99}$ & ---\\
            $\log_{10}\rm CH_4$ & $\mathcal{U}(-12, -2)$ & $-11.30_{-0.50}^{+0.68}$ & ---\\
            $\log_{10}P_{\rm base}$ & $\mathcal{U}(-6, 2)$ & $-3.86_{-0.27}^{+0.30}$ & bar \\
            $\log_{10}\alpha_{\rm cld}$ & $\mathcal{U}(-2, 2)$ & $1.29_{-0.45}^{+0.42}$ & --- \\
            $\log_{10}\kappa_{\rm cld}$ & $\mathcal{U}(-6, 6)$ & $-0.24_{-0.28}^{+0.24}$ & cm$^{2}$\,g$^{-1}$ \\
            $\log_{10}q_{\rm c}$ & $-1$ (fixed) & --- & g g$^{-1}$ \\
            \hline
        \end{tabular*}
    \end{minipage}
\end{table*}

\begin{table*}[t]
    \centering    
    \begin{minipage}[t]{0.48\textwidth}
        \centering
        % Ret70
        \caption{Prior and posterior values for the atmospheric parameters from the 1.5D \texttt{Exo Skryer} equilibrium chemistry retrieval of \planet.}
        \label{tab:chem_eq}
        \vspace{2mm}
        \scriptsize
        \begin{tabular*}{\textwidth}{@{\extracolsep{\fill}}lccc}
            \toprule
            Parameter & Prior & Posterior & Units \\ 
            \hline
            \multicolumn{4}{@{}l}{\textit{Shared}} \\
            $\log_{10}P_{\rm ref}$ & $\mathcal{U}(-8, 2)$ & $1.46_{-0.41}^{+0.35}$ & bar \\
            $\log_{10}g$ & $\mathcal{U}(2.5, 3.5)$ & $2.524_{-0.016}^{+0.027}$ & cm\,s$^{-2}$ \\
            ${\rm F444\rm W}_{\rm offset}$ & $\mathcal{U}(-200, 200)$ & $-11.5_{-20.9}^{+24.3}$ & ppm \\ \\
            
            \multicolumn{4}{@{}l}{\textit{Morning}} \\
            $T_{\rm iso}$ & $\mathcal{U}(500, 3000)$ & $1689_{-75}^{+93}$ & K \\
            $\rm [M/H]$ & $\mathcal{U}(-2, 3)$ & $0.34_{-0.35}^{+0.25}$ & --- \\
            $\rm C/O$ & $\mathcal{U}(0.1, 2)$ & $0.26_{-0.10}^{+0.18}$ & --- \\
            $\log_{10}P_{\rm base}$ & $\mathcal{U}(-6, 2)$ & $-2.21_{-1.62}^{+2.40}$ & bar \\
            $\log_{10}\alpha_{\rm cld}$ & $\mathcal{U}(-2, 2)$ & $0.81_{-0.77}^{+0.55}$ & --- \\
            $\log_{10}\kappa_{\rm cld}$ & $\mathcal{U}(-6, 6)$ & $0.40_{-0.24}^{+0.23}$ & cm$^{2}$\,g$^{-1}$ \\ 
            $\log_{10}q_{\rm c}$ & $-1$ (fixed) & --- & g g$^{-1}$ \\ \\
            
            \multicolumn{4}{@{}l}{\textit{Evening}} \\
            $T_{\rm iso}$ & $\mathcal{U}(500, 3000)$ & $1731_{-81}^{+93}$ & K \\
            $\rm [M/H]$ & $\mathcal{U}(-2, 3)$ & $0.57_{-0.23}^{+0.26}$ & --- \\
            $\rm C/O$ & $\mathcal{U}(0.1, 2)$ & $0.28_{-0.12}^{+0.18}$ & --- \\
            $\log_{10}P_{\rm base}$ & $\mathcal{U}(-6, 2)$ & $-3.07_{-1.47}^{+2.50}$ & bar \\
            $\log_{10}\alpha_{\rm cld}$ & $\mathcal{U}(-2, 2)$ & $1.02_{-0.81}^{+0.55}$ & --- \\
            $\log_{10}\kappa_{\rm cld}$ & $\mathcal{U}(-6, 6)$ & $0.40_{-0.19}^{+0.16}$ & cm$^{2}$\,g$^{-1}$ \\
            $\log_{10}q_{\rm c}$ & $-1$ (fixed) & --- & g g$^{-1}$ \\
            \hline
        \end{tabular*}
    \end{minipage}
    \hfill 
    \begin{minipage}[t]{0.48\textwidth}
        \centering
        \caption{Prior and posterior values for the atmospheric parameters from the 1.5D and 1D \texttt{petitRADTRANS} equilibrium chemistry retrieval of \planet. Note that for planetary mass and radius we use priors that were $\pm5\%$ and $\pm20\%$ from the literature value, respectively, exact values are listed below.}
        \label{tab:prt_priors}       
        \vspace{2mm}
        \scriptsize
        \begin{tabular*}{\textwidth}{@{\extracolsep{\fill}}lccc}
            \toprule
            Parameter & Prior & Posterior & Units \\ 
            \hline
            \multicolumn{4}{@{}l}{\textbf{1.5D retrieval}} \\
            \multicolumn{4}{@{}l}{\textit{Shared}} \\
            Mass & $\mathcal{U}(0.6935, 0.7665)$ & $0.742_{-0.025}^{+0.018}$ & $M_\mathrm{jup}$\\
            Radius & $\mathcal{U}(1.112, 1.668)$ & $1.3929\pm0.0012$ & $R_\mathrm{jup}$ \\
            % ${\rm F444\rm W}_{\rm offset}$ & $\mathcal{U}(-200, 200)$ & $18.6_{-19.3}^{+20.1}$ & ppm \\ 
            $\rm C/O$ & $\mathcal{U}(0.1, 2)$ & $0.240_{-0.091}^{+0.15}$ & --- \\
            $\rm [M/H]$ & $\mathcal{U}(-2, 3)$ & $-0.31_{-0.20}^{+0.24}$ & --- \\ \\
            \multicolumn{4}{@{}l}{\textit{Morning}} \\
            $T_{\rm iso}$ & $\mathcal{U}(500, 3000)$ & $902_{-60}^{+46}$ & K \\
            $\log_{10}P_{\rm cloud}$ & $\mathcal{U}(-6, 2)$ & $-2.6_{-3.4}^{+3.1}$ & bar \\\\
            \multicolumn{4}{@{}l}{\textit{Evening}} \\
            $T_{\rm iso}$ & $\mathcal{U}(500, 3000)$ & $1031_{-70}^{+54}$ & K \\
            $\log_{10}P_{\rm cloud}$ & $\mathcal{U}(-6, 2)$ & $-3.1_{-3.1}^{+3.5}$ & bar \\ \\\\
            \multicolumn{4}{@{}l}{\textbf{1D retrieval}} \\
            Mass & $\mathcal{U}(0.6935, 0.7665)$ & $0.726_{-0.023}^{+0.025}$ & $M_\mathrm{jup}$\\
            Radius & $\mathcal{U}(1.112, 1.668) $ & $1.3654_{-0.0047}^{+0.0049}$ & $R_\mathrm{jup}$ \\
            ${\rm F444\rm W}_{\rm offset}$ & $\mathcal{U}(-200, 200)$ & $4.57_{-23}^{+8.6}$ & ppm \\ 
            ${\rm STIS\ G430}_{\rm offset}$ & $\mathcal{U}(-500, 500)$ & $68_{-53}^{+51}$ & ppm \\ 
            ${\rm STIS\ G750}_{\rm offset}$ & $\mathcal{U}(-500, 500)$ & $103_{-50}^{+100}$ & ppm \\ 
            ${\rm WFC3\ G141}_{\rm offset}$ & $\mathcal{U}(-500, 500)$ & $-2_{-22}^{+125}$ & ppm \\ 
            $\rm C/O$ & $\mathcal{U}(0.1, 1.5)$ & $0.143_{-0.032}^{+0.060}$ & --- \\
            $\rm [M/H]$ & $\mathcal{U}(-1, 2)$ & $0.24\pm 0.14$ & --- \\ 
            $T_{\rm iso}$ & $\mathcal{U}(500, 3000)$ & $1216\pm 64$ & K \\
            $\log_{10}P_{\rm cloud}$ & $\mathcal{U}(-8, 2)$ & $-1.6 \pm 2.4$ & bar \\
            $\kappa_0$ & $\mathcal{U}(-20, 2)$ & $14.6_{-5.4}^{+3.8}$ & cgs \\
            $\gamma$ & $\mathcal{U}(-20, 2)$ & $-1.85_{-2.8}^{+0.25}$ & --- \\
            $\phi$ & $\mathcal{U}(0, 1)$ & $0.706_{-0.042}^{+0.087}$ & --- \\
            \hline
        \end{tabular*}
    \end{minipage}
\end{table*}
\clearpage

\bibliography{ref,references}{}
\bibliographystyle{aasjournalv7}


\end{document}